\documentclass[aps,prx,twocolumn,amsfonts,amsmath,amssymb,footinbib,showpacs,longbibliography,superscriptaddress,10pt]{revtex4-2}

\usepackage[english]{babel} 
\usepackage[utf8]{inputenc} 
\usepackage[T1]{fontenc} 
\usepackage{physics} 
\usepackage{siunitx} 
\usepackage{amsmath} 
\usepackage{amssymb} 
\usepackage{amstext} 
\usepackage{graphicx} 
\usepackage[hidelinks]{hyperref} 
\usepackage[dvipsnames]{xcolor} 
\usepackage{lipsum} 
\usepackage{array} 
\usepackage[all]{hypcap} 
\usepackage[normalem]{ulem}
\usepackage[ruled,nofillcomment,linesnumbered]{algorithm2e} 

\usepackage{latexsym}
\usepackage{subfigure}
\usepackage{epsfig}

\hypersetup{
    colorlinks=true,
    breaklinks=true,
    urlcolor=black,
    linkcolor=NavyBlue,
    citecolor=NavyBlue,
    bookmarksopen=true,
    pdftoolbar=false,
    pdfmenubar=false,
}

\newcommand{\ie}{i.e.}
\newcommand{\eg}{e.g.}

\newcommand{\e}{\textrm{e}} 
\newcommand{\hc}{\text{h.c.}} 
\renewcommand{\a}{\Hat{a}} 
\newcommand{\adag}{\Hat{a}^\dagger} 
\newcommand{\adagd}{\Hat{a}^{\dagger 2}} 
\newcommand{\ta}{\tilde{a}} 
\newcommand{\tadag}{\tilde{a}^\dagger} 
\newcommand{\tadagd}{\tilde{a}{^\dagger 2}} 
\renewcommand{\b}{\Hat{b}} 
\newcommand{\bdag}{\Hat{b}^\dagger} 
\renewcommand{\H}{\Hat{H}} 
\newcommand{\hrho}{\Hat{\rho}} 
\newcommand{\trho}{\tilde{\rho}} 
\newcommand{\hphi}{\Hat{\varphi}} 
\newcommand{\rhol}{\hrho_l} 
\newcommand{\LD}{\Hat{L}_2} 
\newcommand{\dm}[2]{\ket{#1}\hspace{-2px}\bra{#2}}

\colorlet{mmcolor}{orange!100!}

\begin{document}

\title{
Combined Dissipative and Hamiltonian Confinement of Cat Qubits
}

\author{Ronan Gautier}
\email{ronan.gautier@inria.fr}
\affiliation{Laboratoire de Physique de l'\'Ecole Normale Sup\'erieure, Inria, ENS, Mines ParisTech, Universit\'e PSL, Sorbonne Universit\'e, Paris, France}

\author{Alain Sarlette}
\affiliation{Laboratoire de Physique de l'\'Ecole Normale Sup\'erieure, Inria, ENS, Mines ParisTech, Universit\'e PSL, Sorbonne Universit\'e, Paris, France}
\affiliation{Department of Electronics and Information Systems, Ghent University, Belgium}

\author{Mazyar Mirrahimi}
\affiliation{Laboratoire de Physique de l'\'Ecole Normale Sup\'erieure, Inria, ENS, Mines ParisTech, Universit\'e PSL, Sorbonne Universit\'e, Paris, France}


\begin{abstract}
Quantum error correction with biased-noise qubits can drastically reduce the hardware overhead for universal and fault-tolerant quantum computation. Cat qubits are a promising realization of biased-noise qubits as they feature an exponential error bias inherited from their non-local encoding in the phase space of a quantum harmonic oscillator. To confine the state of an oscillator to the cat qubit manifold, two main approaches have been considered so far: a Kerr-based Hamiltonian confinement with high gate performances, and a dissipative confinement with robust protection against a broad range of noise mechanisms. We introduce a new combined dissipative and Hamiltonian confinement scheme based on two-photon dissipation together with a Two-Photon Exchange (TPE) Hamiltonian. The TPE Hamiltonian is similar to Kerr nonlinearity, but unlike the Kerr it only induces a bounded distinction between even- and odd-photon eigenstates, a highly beneficial feature for protecting the cat qubits with dissipative mechanisms. Using this combined confinement scheme, we demonstrate fast and bias-preserving gates with drastically improved performance compared to dissipative or Hamiltonian schemes. In addition, this combined scheme can be implemented experimentally with only minor modifications of existing dissipative cat qubit experiments.
\end{abstract}

\maketitle


\section{Introduction}\label{sec:intro}
Superconducting qubits are promising candidates for realizing a first practical quantum computer. From the first Cooper pair box~\cite{nakamura1999coherent} to the acclaimed transmon~\cite{koch2007charge}, and more recently with qubits such as the fluxonium~\cite{manucharyan2009fluxonium} or the 0-$\pi$ qubit~\cite{gyenis2021experimental}, superconducting circuits have been steadily improving in the past 20 years. Nevertheless, there are still orders of magnitude in error mitigation until the first fault tolerant computers can be attained. Using quantum error correction (QEC), it is possible to engineer high-quality logical qubits by delocalizing quantum information in arrays of lower-quality physical qubits, provided that the latter feature low enough error rates to reach fault-tolerance thresholds~\cite{shor1996fault,fowler2012surface,campbell2017roads}. Engineering physical qubits with low error rates, both when idling and during quantum gates, is thus an essential task that needs to be overcome if we hope to reach fault-tolerance.

Recently, error-biased qubits have been actively investigated for their promise of hardware-efficient and fault-tolerant universal quantum computation~\cite{aliferis2008fault,aliferis2009fault, tuckett2018ultrahigh,tuckett2019tailoring,huang2020fault,bonilla2021xzzx,higgott2021subsytem}. In particular, the bosonic cat qubit encoding~\cite{mirrahimi2014dynamically} is showing potential thanks to its exponential error bias~\cite{lescanne2020exponential} inherited from a non-local encoding in the phase space of a harmonic oscillator. In the context of QEC, such a qubit with exponential error bias can drastically reduce hardware overheads by encoding information in elongated surface codes~\cite{chamberland2020building}, in XZZX surface codes~\cite{darmawan2021practical}, or more simply in one-dimensional repetition codes~\cite{guillaud2019repetition, guillaud2021error}. The latter situation can be viewed as an effective Bacon-Shor type concatenated code~\cite{bacon2006operator, shor1995scheme}, but where the cat qubit condenses the inner code into a single physical element, with inner code distance corresponding to the harmonic oscillator mean number of photons. The inner code correction circuit, in this viewpoint, must be replaced by a mechanism confining the state of the quantum harmonic oscillator to the two-dimensional manifold hosting the cat qubit.

In the recent years two approaches have been proposed in order to achieve this confinement. The first approach is based on engineered two-photon dissipation which attracts any initial state towards the cat qubit manifold~\cite{mirrahimi2014dynamically,leghtas2015confining,lescanne2020exponential}. The second approach is a Hamiltonian confinement based on two-photon driving and Kerr non-linearity~\cite{puri2017engineering,grimm2020stabilization}. The manifold associated to the cat qubit states corresponds to a degenerate eigenspace of the engineered Hamiltonian separated from the rest of the spectrum with a gap proportional to the strength of the Kerr non-linearity.

The first confinement approach benefits from the advantage that the states in the cat qubit manifold correspond to the only steady states of the dissipative mechanism which therefore counteracts any potential leakage outside the code space. Thus any undesired perturbation to the ideal dynamics, such as the usual decoherence effects, lead only to bit-flip and/or phase-flip errors in the code space. Furthermore, as discussed in~\cite{cohen2017autonomous} and experimentally demonstrated in~\cite{lescanne2020exponential}, the associated bit-flip errors are exponentially suppressed with the oscillator mean number of photons. This bit-flip suppression can replace the inner code of a Bacon-Shor type concatenated code, as mentioned above, paving the way towards a hardware-efficient scaling for fault-tolerance~\cite{guillaud2019repetition,guillaud2021error,chamberland2020building}. The main inconvenience of this approach is currently the limited performance of the logical gates. As detailed in~\cite{mirrahimi2014dynamically,guillaud2019repetition}, various bias-preserving gates such as single qubit rotations $Z_\theta$, two-qubit CNOT and three-qubit Toffoli gates can be performed through adiabatic processes. Most importantly, the bit-flip errors remain exponentially suppressed throughout these processes~\cite{touzard2018coherent, guillaud2019repetition,chamberland2020building}. However, these processes do lead to significant phase-flip errors, causing an additional burden for their suppression through a repetition code. In practice, this leads to challenging requirements for the ratio between the engineered two-photon dissipation rate and the undesired relaxation and decoherence rates to reach the fault-tolerance threshold of the outer repetition code.
 
On the other hand, in the Kerr-based confinement approach, gate performance can be improved through the application of superadiabatic pulse designs~\cite{xu2021engineering}, which take advantage of the purely Hamiltonian evolution to suppress gate-induced leakage out of the code space. However, in the absence of dissipative stabilization of the cat qubit manifold, leakage induced by perturbations different from quasi-static Hamiltonians such as thermal excitation or photon dephasing, is not countered. As shown in~\cite{putterman2021colored} and further argued in this paper, this leakage can in turn lead to significant bit-flip errors that suppress the noise bias and therefore block the way towards hardware-efficient fault-tolerance. Recently, the authors of~\cite{putterman2021colored} have proposed a promising approach to remedy this problem and to ensure the suppression of bit-flips. This approach consists in the addition of a colored relaxation to compensate for the leakage out of the cat manifold. Through this addition, the bit-flip suppression would be re-established for the Kerr cat, but to reach the same level of performance as in the case of the two-photon dissipation, one would need a careful hardware engineering of the bath beyond the Purcell filters that are routinely used in superconducting devices.

It is tempting to think that by combining the two mechanisms of two-photon dissipation and Kerr Hamiltonian confinement, we might benefit from the best of both worlds. This idea is valid but it comes with some important limitations. First, the realization of the bias-preserving operations would require the combination of the engineered Hamiltonians and dissipators as laid out in~\cite{guillaud2019repetition} and~\cite{puri2020bias}. This is a daunting experimental task which could lead to unanticipated roadblocks. Second, as it will become clear in the analysis of this paper, in order to re-establish a similar level of bit-flip suppression to the purely dissipative mechanism, the Kerr strength cannot surpass the two-photon decay rate. This significantly reduces the interest of the Kerr confinement in performing fast and high-fidelity gates.

To overcome this difficulty, a main contribution of the present paper is to introduce an alternative cat qubit confinement Hamiltonian that does not suffer from the  limitations of the Kerr confinement, and which can be combined with a dissipative protection mechanism. Note that while this paper focuses on bosonic cat qubits, the idea is rather general. Indeed, the protection of quantum information generally relies on either a Hamiltonian gap, where the information is encoded in a degenerate eigenspace of an engineered Hamiltonian, or a dissipative mechanism, usually based on reservoir engineering or feedback control. Our paper shows that the combination of these two techniques is not straightforward. More precisely, for such combinations one cannot simply rely on the energy gap but it is also needed to carefully engineer the spectrum of excited energy levels. We therefore hope that similar methods could be replicated on other quantum information devices.

The purpose of the present paper is thus to propose a highly efficient scheme for combined Hamiltonian and dissipative cat qubit confinement, and to analyze its performance compared to other schemes. Section~\ref{sec:bitflip-confinement} first investigates the exponential suppression of bit-flip errors for two existing confinement schemes: two-photon dissipation and Kerr confinement. While the former features a reliable exponential error-bias, the latter is susceptible to both thermal and dephasing noise which cause leakage out of the cat qubit codespace and in turn considerable bit-flip errors. To address this difficulty, the combined Kerr and two-photon dissipation confinement is studied. The Kerr spectrum, beyond the degenerate ground states spanning the cat qubit, features diverging energy level spacing between consecutive eigenstates. We show how this limits the advantages of this combined confinement.

Section~\ref{sec:tpe} introduces the Two-Photon Exchange (TPE) Hamiltonian, a nonlinear coupling between a resonator and a two-level buffer system. The TPE Hamiltonian provides a similar confinement to the Kerr Hamiltonian, but its spectrum features uniformly bounded energy level spacing between consecutive eigenstates. This essential feature enables an efficient combination with two-photon dissipation for protecting the cat qubit manifold from noise-induced leakage. In addition, such combined TPE and two-photon dissipation scheme is readily implementable experimentally with only minor modifications of the setup implemented in~\cite{lescanne2020exponential}. A quantitative analysis further yields working points for the ratio of Hamiltonian to dissipative rates, both for Kerr and TPE combined confinement schemes, to retain a desired exponential suppression of bit-flip errors.

In Section~\ref{sec:zgate}, the performance of a single-qubit Z gate for the combined TPE confinement is evaluated at the identified working point and compared with existing bias-preserving schemes. Using the combined TPE scheme, gate times can be drastically reduced and gate fidelities improved by a factor of up to 400 compared to fully dissipative gate implementations, and so without compromising bit-flip errors. Indeed, the combined confinement scheme maintains the protection against broadband noise both when idling and during gates unlike fully Hamiltonian schemes. In Section~\ref{sec:cnot}, the performance of two-qubit CNOT gates is investigated. Similarly, important improvements in gate speeds and gate fidelities are shown compared to existing bias-preserving schemes. Here, the CNOT design that we study finds its roots in~\cite{puri2019stabilized} and benefits from a simpler implementation than~\cite{guillaud2019repetition} and~\cite{puri2020bias}.

\begin{figure*}[t!]
         \centering
         \includegraphics[width = 2.05\columnwidth]{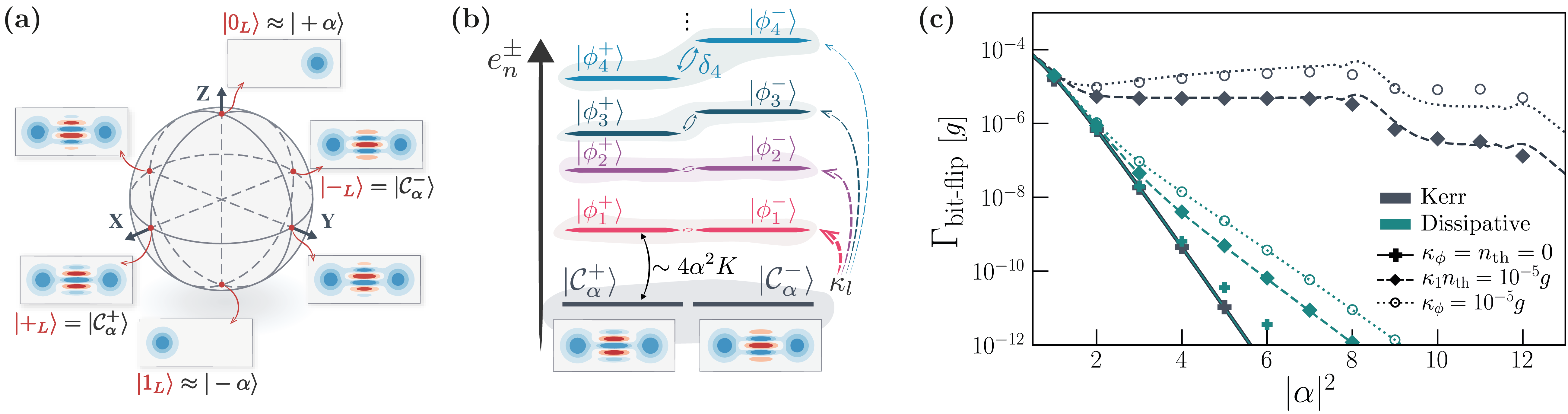}
\caption{\label{fig:kerr-intro}
(a)~Bloch sphere representation of a cat qubit. (b)~Energy spectrum of the Kerr Hamiltonian separated in two branches of even and odd parity eigenstates. The ground eigenspace is the cat qubit manifold (codespace), gapped from excited eigenstates by an energy $\lvert e_1^\pm - e_0 \rvert \approx 4 |\alpha|^2 K$. Leakage out of the codespace at rate $\kappa_l = \kappa_1 n_\text{th} + |\alpha|^2 \kappa_\phi$ readily brings the system towards a superposition of all excited states $\ket{\phi_{n}^{\pm}}$, where even and odd excited eigenstates dephase at rate $\delta_n = |e_n^- - e_n^+|$. (c)~Bit-flip error rate of an idle cat qubit confined with a Kerr Hamiltonian (grey, $g \equiv K$) or with two-photon dissipation (green, $g \equiv \kappa_2$), in units of $g$. The cat qubit is subject to single-photon loss with rate $\kappa_1 = 10^{-3} g$ (crosses, solid lines), with additional thermal noise $\kappa_1 n_\text{th} = 10^{-5}g$ (diamonds, dashed lines) or pure dephasing $\kappa_\phi = 10^{-5}g$ (circles, dotted lines). Markers indicate numerical data, lines indicate analytical fits.
}
 \end{figure*}

Finally, Section~\ref{sec:exp} presents a proposal for the experimental realization of a combined TPE and two-photon dissipation confinement scheme. This proposal is based on the dissipative cat qubit implementation of~\cite{lescanne2020exponential} and makes use of an Asymmetrically Threaded SQUID (ATS). We also investigate the influence of various experimental imperfections on the scheme's performance.
Overall, the combined TPE and dissipative scheme offers the same noise protection as with purely dissipative schemes, but with important improvements in cat qubit gate speeds as well as additional protection from spurious Hamiltonians with only minimal hardware overhead.


\section{Cat confinement and bit-flip suppression}\label{sec:bitflip-confinement}

Cat qubits are error biased qubits encoded in quantum harmonic oscillators as the two-dimensional subspace spanned by two-component Schr\"{o}dinger cat states~\cite{cochrane1999macroscopically,ralph2003quantum,mirrahimi2014dynamically}. The cat qubit logical basis is defined as
\begin{equation}
    \begin{split}
        \ket{0_L} =&~\frac{1}{\sqrt{2}} \left(\ket{\mathcal{C}_\alpha^+} + \ket{\mathcal{C}_\alpha^-} \right) = \ket{\alpha} + \mathcal{O}(e^{-2|\alpha|^2}) \\ 
        \ket{1_L} =&~\frac{1}{\sqrt{2}} \left(\ket{\mathcal{C}_\alpha^+} - \ket{\mathcal{C}_\alpha^-} \right) = \ket{-\alpha} + \mathcal{O}(e^{-2|\alpha|^2})
    \end{split}
\end{equation}
where $\ket{\mathcal{C}_\alpha^\pm}=(\ket{\alpha}\pm\ket{-\alpha}) / \mathcal{N}_{\pm}$ are the even and odd parity cat states, and $\mathcal{N}_{\pm}$ normalization constants. A typical Bloch sphere representation of a cat qubit is shown in Figure~\ref{fig:kerr-intro}(a). The error bias of cat qubits is inherited from their non-local encoding in phase space. It is shown that their typical bit-flip error rate is exponentially suppressed in the cat qubit mean number of photons, with only a linear increase in phase error rates~\cite{cohen2017autonomous,lescanne2020exponential}. We refer to the Appendices~\ref{sec-apdx:catqubitdef} and~\ref{sec-apdx:bitflipest} for further details on the cat qubit definition and the estimation of bit-flip and phase-flip error rates.

Let us consider a quantum harmonic oscillator under some cat qubit confinement scheme and typical decoherence effects. The master equation governing the evolution of the oscillator is given by
\begin{equation}\label{eq:mastereq}
    \frac{\dd \hrho}{\dd t} = g \mathcal{L}_\text{conf}\hrho + \kappa_- \mathcal{D}[\a]\hrho + \kappa_+ \mathcal{D}[\adag]\hrho + \kappa_\phi \mathcal{D}[\adag \a]\hrho \, .
\end{equation}
Here $\mathcal{D}[\Hat{L}] \hrho = \Hat{L} \hrho \Hat{L}^\dagger - \{\Hat{L}^\dagger \Hat{L}, \hrho\} / 2$ is the dissipation super-operator associated to channel $\Hat{L}$, applied with pure dephasing rate $\kappa_\phi$, single-photon relaxation and excitation rates $\kappa_-=\kappa_1 (1+n_\text{th})$ and $\kappa_+ = \kappa_1 n_\text{th}$, with $n_\text{th}$ the average number of thermal photons. The superoperator $\mathcal{L}_\text{conf}$ denotes a confinement scheme, \ie~a super-operator with a two-dimensional steady state manifold equal to the cat qubit codespace.

For dissipative cat qubits, the confinement scheme takes the form of a two-photon driven dissipation $\mathcal{L}_\text{conf} = \mathcal{D}[\LD]$ with $\LD = \a^2 - \alpha^2$ and where the bare confinement rate is the two-photon dissipation rate $g \equiv \kappa_2$~\cite{mirrahimi2014dynamically}. This confinement scheme is also an asymptotic stabilization scheme in the sense of dynamical systems theory, meaning that any initial state will asymptotically converge to the cat qubit manifold. The stabilization is furthermore exponential since any initial state in a close neighborhood of the cat qubit manifold will converge exponentially fast to a steady state in this manifold. In the limit $\kappa_2 \gg \kappa_1$, the typical time scale of this convergence is given by a re-scaled confinement rate $\kappa_\text{conf} = 4 |\alpha|^2 \kappa_2$~\cite{azouit2016well}.

For Kerr cat qubits, the confinement scheme is instead an energy-conserving superoperator $\mathcal{L}_\text{conf} = i [\H_\text{Kerr}, \cdot]$ where the Kerr Hamiltonian reads $\H_\text{Kerr} = (\adagd - \alpha^{* 2}) (\a^2 - \alpha^2)$ and with confinement rate given by the Kerr nonlinearity $g \equiv K$~\cite{puri2017engineering}. This Hamiltonian features an exactly degenerate ground eigenspace given by the cat qubit manifold, which is gapped from other eigenstates with an energy approximately given by $4|\alpha|^2 K$ in the limit of large $\alpha$, as shown in Figure~\ref{fig:kerr-intro}(a). This energy gap protects the qubit from spurious weak and slowly varying Hamiltonian perturbations, as the perturbed eigenspace will remain very close to the cat qubit manifold. However, when the state leaks out, no process ensures its asymptotic convergence back to the cat qubit manifold, and we therefore favor the terminology of ``confinement'' rather than ``stabilization''. The excited eigenstates of the Kerr Hamiltonian can be separated into two branches of even and odd photon number parities, which asymptotically converge towards $\ket{\mathcal{C}_\alpha^+}$ and $\ket{\mathcal{C}_\alpha^-}$ respectively under two-photon dissipation. Since modifying the phase between even and odd cat states corresponds to bit-flips, we would ideally want these two excited branches to be degenerate like the cat states. However, the Kerr Hamiltonian excited eigenstates are all non-degenerate, with energy level spacing between consecutive energy levels increasing with the excitation number, as shown on Fig.~\ref{fig:kerr-intro}(b). Although each spacing can be suppressed exponentially with the mean number of photons $|\alpha|^2$, this suppression only kicks in at $|\alpha|^2 \gtrsim 4n$ for the $n$-th pair of excited states~\cite{puri2017engineering,putterman2021colored}. As discussed in~\cite{putterman2021colored} and further argued in this paper, this fact has major consequences on the effective bit-flip suppression for such a confinement.

In Figure~\ref{fig:kerr-intro}(c),  the bit-flip error rates of idling dissipative cat qubits and idling Kerr cat qubits are presented in the presence of various noise processes. While markers show simulation data at large idling times, lines show analytical predictions as detailed in Appendix~\ref{sec-apdx:bitflipest}. For $\kappa_\phi = n_\text{th} = 0$, both schemes feature an exponential bit-flip error suppression, proportional to $\kappa_1 |\alpha|^2 \exp(-4 |\alpha|^2)$. In the presence of thermal or dephasing noise, dissipative cat qubits retain an exponential error bias with exponential suppression proportional to 
$\kappa_l \exp(-2|\alpha|^2)$ where $\kappa_l = \kappa_1 n_\text{th} + |\alpha|^2 \kappa_\phi$ is the leakage rate out of the cat qubit manifold. Kerr cat qubits on the other hand feature an approximately constant bit-flip error rate given by the leakage rate $\kappa_l$. The predictions fit simulation data very closely, and we now detail the physical process of Kerr-induced bit-flip errors behind the models used here.

The process is the following. Thermal and dephasing noise induce leakage out of the codespace at a rate denoted by $\kappa_l$, which is \emph{a priori} small compared to the Kerr nonlinearity, $\kappa_l \ll K$. This leakage is slow but, in the absence of any mechanism attracting it back to the cat qubit manifold, over a time $t$ it deforms a state initially in the cat qubit manifold into a new state that is about $\kappa_l t$-removed from the steady state manifold. This new state has a finite overlap with all excited eigenstates of the Kerr Hamiltonian (denoted as $\ket{\phi_n^\pm}$) which form an orthogonal basis of the Hilbert space. Let us consider the $n$-th pair of Kerr eigenstates. These two eigenstates dephase at a rate given by their energy level spacing $\delta_n = e_n^- - e_n^+$. Since they are by definition eigenstates of even and odd parities, some dephasing of phase angle $\pi$ will result in an effective bit-flip error. With respect to the above discussion, single-photon loss is in fact a confinement mechanism attracting the leaked state back to the cat-qubit manifold at the rate $\kappa_1$. However, as $\kappa_1\ll K$, the system will potentially undergo bit-flip errors before this reconvergence to the code space. In the Appendix~\ref{sec-apdx:bitflipest}, this model is detailed and, over a time $t$ short compared to rate of leakage, $\kappa_l t \ll 1$, and short compared to the rate of reconvergence, $\kappa_1 t \ll 1$, it yields the following bit-flip error probability,
\begin{equation}\label{eq:kerr-bitflip}
    p_X(t) = \kappa_{l} t \sum_{n > 0} \lambda_n \left[ 1 - \frac{\sin(\delta_n t)}{\delta_n t} \right] \; .
\end{equation}
Here $\lambda_n = \sum_\pm |\braket{\phi_n^\pm | \alpha, 1}|^2 / 2$ where $\ket{\alpha,1}$ is the first displaced Fock state $\hat{D}(\alpha) \ket{\hat{n}=1}$, \ie~the state obtained at first order under thermal excitation or photon dephasing when starting at $\ket{\alpha}$ in the cat qubit codespace. In this sense, $\kappa_l \lambda_n$ represents the rate at which each pair of Kerr excited eigenstates is populated by codespace leakage. In addition, the right-hand side bracket of Eq.~\eqref{eq:kerr-bitflip} represents how the dephasing of the even and odd parity eigenstates, accumulated over time $t$ before getting back to the codespace, translates to bit-flip errors. This bracket vanishes for small time or energy level spacing $\delta_n t \ll 1 $ and equals $1$ for large time or energy level spacing $\delta_n t\gg 1$, as expected. 
 \begin{figure}[t!]
         \centering
         \includegraphics[width = 1.0\columnwidth]{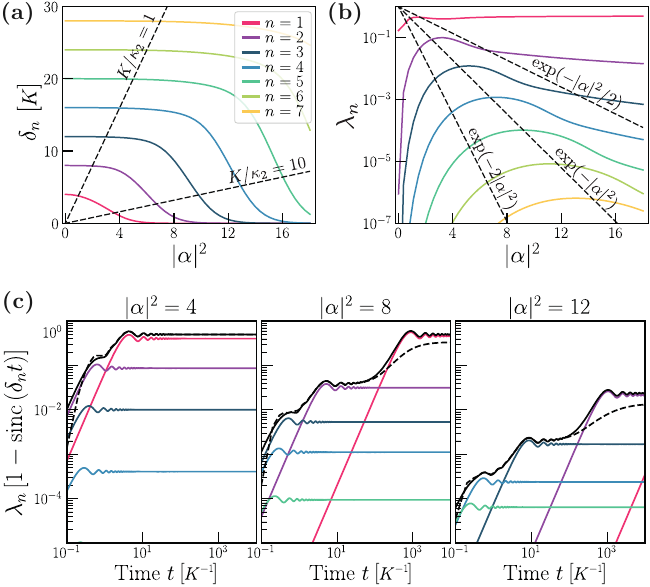}
\caption{\label{fig:kerr-degeneracies}
(a)~Energy level spacing between even and odd parity eigenstates of the Kerr Hamiltonian $\delta_n = |e_n^- - e_n^+|$ in units of $K$. Dashed lines represent the threshold $\delta_n = \kappa_\text{conf}$ at different values of $K / \kappa_2$. Only eigenstates below this threshold benefit from the additional two-photon dissipation scheme, meaning that $\delta_n / \kappa_\text{conf} \ll 1$ in Eq.~\eqref{eq:kerr-gammabitflip-est}. (b)~Relative overlap between the $n$-th pair of excited Kerr eigenstates and a cat qubit state deformed by thermal or dephasing noise (cf. Appendix~\ref{sec-apdx:bitflipest}). Dashed lines represent the threshold $\kappa_l \lambda_n = \kappa_l \exp(-c |\alpha|^2)$, which one may want to impose with certain values of $c$ in order to achieve exponential suppression of bit-flip errors. All eigenstates above the threshold are susceptible to contribute to a bit-flip rate larger than the target rate via Eq.~\eqref{eq:kerr-gammabitflip-est}. (c)~Individual terms of the sum of Eq.~\eqref{eq:kerr-bitflip} with respect to time. The solid black line represents the sum of all terms. The dashed black line represents $p_X(t) / \kappa_l t$ extracted from a numerical simulation of Kerr confinement with single-photon loss $\kappa_1 = 10^{-3} K$ and thermal noise $n_\text{th} = 10^{-2}$.
}
\end{figure}

With this we can put some numbers on bit-flip error probabilities under Kerr confinement by looking at Fig.~\ref{fig:kerr-degeneracies} in which panels (a) and (b) show respectively $\delta_n$ and $\lambda_n$ with respect to $|\alpha|^2$ for the first pairs of Kerr eigenstates, while panel (c) shows each term of Eq.~\eqref{eq:kerr-bitflip} with respect to time for $|\alpha|^2 = 4$, $8$ and $12$. Please discard the black dashed lines on panels (a) and (b) for the moment. Together, these figures show how the various eigenstates contribute to bit-flip errors via Eq.~\eqref{eq:kerr-bitflip} for each mean number of photons and each time. 
Note thus that according to Fig.~\ref{fig:kerr-degeneracies}(b), a distribution over all Kerr eigenstates is induced by a single thermal photon or dephasing perturbation. In turn, the associated $\delta_n$ increases with $n$ and features two regimes as a function of $\alpha$ (constant, then exponentially suppressed) as shown on Fig.~\ref{fig:kerr-degeneracies}(a); a log-scale version of the latter  is also shown in appendix for better readability. The overall bit-flip rate results from the combination of these two effects.

For instance, at $|\alpha|^2 = 12$ and considering a typical time $t \sim 10 / K$, Fig.~\ref{fig:kerr-degeneracies}(c) shows that the $n=3$ pair of excited Kerr eigenstates contributes the most to the sum of Eq.~\eqref{eq:kerr-bitflip}. Indeed, we find that $\delta_3 t \approx 5.2$, thus contributing to a bit-flip error dominated by $\kappa_l t \lambda_3$ with $\lambda_3 \approx 1.7 \cdot 10^{-3}$. In comparison, the dissipative cat qubit error probability is also proportional to $\kappa_l t$ but with an exponential prefactor $\exp(-2 |\alpha|^2) \approx 4 \cdot 10^{-11}$. Terms with lower $n$ indices contribute less to Kerr-induced bit-flip errors because the $\delta_n$ are well inside the exponentially suppressed regime, \eg~for the second pair of excited states,~$1 - \mathrm{sinc}(\delta_2 t) \approx 2.5 \cdot 10^{-4}$. On the other hand, terms with larger $n$ indices contribute with $1- \mathrm{sinc}(\delta_n t) \approx 1$ even at shorter time scales, but with smaller effect since $\lambda_n$ decreases exponentially with $n$~\cite{putterman2021colored}. The study of \cite{putterman2021colored} shows that at very large values of $|\alpha|^2$ (well above the average photon numbers considered here) one can hope to retrieve an approximate exponential suppression given by $p_X \propto \exp(-0.94 |\alpha|^2)$ which is still less significant than the prefactor $\exp(-2 |\alpha|^2)$ achieved for dissipative cat qubits and with all values of $|\alpha|^2$. Note furthermore that, working at very large values of $|\alpha|^2$ induces significant phase-flip errors and can lead to several other unanticipated difficulties~\cite{chamberland2020building}.

In order to better protect a Kerr cat qubit against thermal and dephasing noise while still benefiting from its excellent performance for gate engineering~\cite{xu2021engineering}, it is interesting to consider combining it with a dissipative confinement scheme. In this paper, we only consider using a two-photon dissipation scheme, but our conclusions hold also for other schemes such as the colored dissipation  introduced in \cite{putterman2021colored}. 

In order to recover the exponential suppression of bit-flip errors, the dissipative stabilization rate should be larger than any significant Kerr-induced dephasing between odd and even photon number subspaces. If this is verified, then leakage out of the cat qubit manifold will reconverge faster towards its initial state thanks to dissipation than it will dephase towards the other side of phase space due to Kerr effects. To make this argument clearer, let us assume a target exponential suppression of bit-flips given by $\Gamma_\text{bit-flip} \propto \exp(-|\alpha|^2)$. Then all eigenstates with $\lambda_n$ above the corresponding dashed line on Fig.~\ref{fig:kerr-degeneracies}(b) could contribute to more bit-flip errors than the targeted exponential. To avoid these contributions, those eigenstates should dephase at a rate $\delta_n$ smaller than the dissipative confinement rate used to reconverge the state. For instance, at $|\alpha|^2 = 12$, the $n\leq5$ eigenstates are above the line $\exp(-|\alpha|^2)$, and their largest degeneracy is $\delta_5 \approx 20 K$. Therefore, the dissipative confinement rate should be $\kappa_\text{conf} = 4 |\alpha|^2 \kappa_2 \gg 20 K$, or equivalently, $\kappa_2 \gg 0.42 K$. This reasoning holds for any value of $\alpha$ and any target exponential rate. One should however bear in mind that two-photon dissipation will induce bit-flip errors proportional to $\kappa_l \exp(-2|\alpha|^2)$ on its own, such that lower error targets cannot be achieved. This explains how to set a lower bound on the ratio $\kappa_2/K$.

Conversely, we may start with an upper bound on $\kappa_2/K$, \eg~from the motivation of sufficiently maintaining the advantages of Kerr Hamiltonian confinement towards implementing gates. Consider for instance that we limit ourselves to $\kappa_2 \leq 0.1 K$. Then Kerr-induced dephasing will be avoided only for the eigenstates below the dashed line $K/\kappa_2=10$ on Fig.~\ref{fig:kerr-degeneracies}(a). At $|\alpha|^2=12$, these states read $n \leq 3$. Eigenstates with $n>3$ will each contribute to bit-flip errors with a rate $\kappa_l \lambda_n$. Fig.~\ref{fig:kerr-degeneracies}(b) indicates that $\lambda_4 \approx 2 \cdot 10^{-4}$, so we have to expect a bit-flip rate $\Gamma_\text{bit-flip} \gtrsim 2 \cdot 10^{-4} \kappa_l$.

 \begin{figure}[t!]
         \centering
         \includegraphics[width = 1.0\columnwidth]{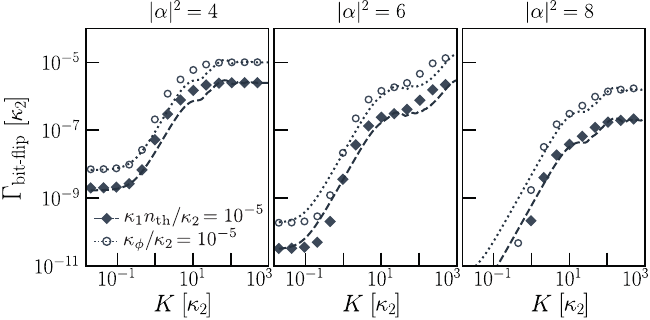}
\caption{\label{fig:kerr-diss-combined}
Bit-flip error rate of an idle cat qubit confined by a combined Kerr and two-photon dissipation for increasing cat sizes. Both axes are graduated in units of $\kappa_2$. The cat qubit is subject to single-photon loss of amplitude $\kappa_1  = 10^{-3}\kappa_2$, with additional thermal excitation $\kappa_1 n_{\text{th}} = 10^{-5} \kappa_2$~(diamonds, dashed lines) or pure dephasing $\kappa_\phi = 10^{-5}\kappa_2 $~(circles, dotted lines). At low enough $K$, the exponential bit-flip suppression is restored. Markers indicate numerical simulation data, lines represent Eq.~\eqref{eq:kerr-gammabitflip-est}.
}
 \end{figure}

Figure~\ref{fig:kerr-diss-combined} shows the actual bit-flip rate when combining Kerr and dissipative confinement schemes. In these simulations, the cat qubit is idling and the amplitude of the Kerr Hamiltonian is varied. The resonator is subject to single-photon loss at rate $\kappa_1  = 10^{-3}\kappa_2$ with either thermal excitations at rate $\kappa_1 n_\text{th}  = 10^{-5}\kappa_2$ (diamonds) or pure dephasing at rate $\kappa_\phi  = 10^{-5}\kappa_2$ (circles). The noise rates considered here and in the rest of this work are within one order of magnitude of already achieved noise rates, both for Kerr~\cite{grimm2020stabilization} and dissipative cat qubits~\cite{touzard2018coherent}, and are expected to be in a feasible range for the next generation of cat qubit experiments. In this Figure, each plot shows a different number of photons $|\alpha|^2 = 4, 6, 8$. In the $K / \kappa_2 \ll 1$ regime, the dissipative cat qubit regime is retrieved with a clear exponential suppression at rate $\exp(-2 |\alpha|^2)$. In the opposite regime of $K / \kappa_2 \gg 1$, the Kerr cat qubit regime is retrieved. Between these two regimes, a smooth transition occurs. The analytical predictions represented with dashed and dotted lines in this figure follow directly from Eq.~\eqref{eq:kerr-bitflip} evaluated at the Kerr excited level characteristic lifetime $t=1/\kappa_{\text{conf}}$, giving
\begin{equation}\label{eq:kerr-gammabitflip-est}
    \begin{split}
        \Gamma_\text{bit-flip} =&~\kappa_1 |\alpha|^2 \exp\left(-4|\alpha|^2\right) \\
        &+ \kappa_l \exp \left(-2|\alpha|^2 \right) \\
        &+ \kappa_l \sum_{n > 0} \lambda_n \left[ 1 - \frac{\sin(\delta_n / \kappa_\text{conf})}{\delta_n / \kappa_\text{conf}} \right].
    \end{split}
\end{equation}
The first two terms are the contribution of two-photon dissipation, whose induced bit-flip errors are exponentially suppressed in $|\alpha|^2$, while the third term is inherited from the trade-off between Kerr dephasing and dissipative confinement. 

The purely dissipative stabilization scheme is ideal for an idling qubit under Markovian noise, as shown here, but Hamiltonian confinement will show its benefits as we add gates or small Hamiltonian perturbations. From this perspective, an ideal working point for the combined confinement scheme would be the largest value of $K$ for which the dissipative exponential suppression remains intact. For the $|\alpha|^2$ values shown here and as further argued in the following section, this corresponds approximately to $K / \kappa_2 = 0.3$. 
Unfortunately, at this working point the Kerr non-linearity turns out to be too small to provide any practical advantage over a purely dissipative scheme. As we have tried to explain, the deterioration of bit-flip protection for larger $K$ values is partly inherited from the increasing energy level spacing $\delta_n$ between even and odd photon number eigenstates of the Kerr Hamiltonian. Indeed, to avoid bit-flips, the dissipative stabilization scheme has to counter the increasing dephasing rates associated to $\delta_n$, for any eigenstate $n$ having significant overlap with a slightly deformed/displaced cat state.

In the following section, we introduce a new cat qubit confinement Hamiltonian that is Kerr-like, but that features bounded energy level spacing between even and odd photon number eigenstates. We observe that this new confinement scheme is indeed much more compatible with a dissipative stabilization. It is furthermore very similar in experimental implementation to two-photon dissipation, thus opening the door to a natural combination of Hamiltonian and dissipative cat qubit confinement.

\section{Two-Photon Exchange Hamiltonian Confinement}\label{sec:tpe}

Two-photon dissipation can be engineered by coupling the cat qubit resonator with a two-photon exchange Hamiltonian $g_2(\a^2 \bdag + \hc)$ to a low-Q buffer mode, that is a resonator $\b$ undergoing strong single-photon loss~\cite{mirrahimi2014dynamically,leghtas2015confining,lescanne2020exponential}. When single-photon relaxation $\kappa_b$ of the buffer mode is large compared to the two-photon exchange rate, the buffer mode can be adiabatically eliminated thus resulting in two-photon dissipation $\kappa_2 \mathcal{D}[\a^2]$ on the cat qubit mode, with rate $\kappa_2 = 4g_2^2 / \kappa_b$~\cite{azouit2017towards, mirrahimi2014dynamically}. To engineer the additional $-\alpha^2$ term of the two-photon dissipator $\LD$, one can drive the buffer mode at its resonance leading effectively to a two-photon drive on the cat qubit mode.

 \begin{figure}[t!]
         \centering
         \includegraphics[width = 1.0\columnwidth]{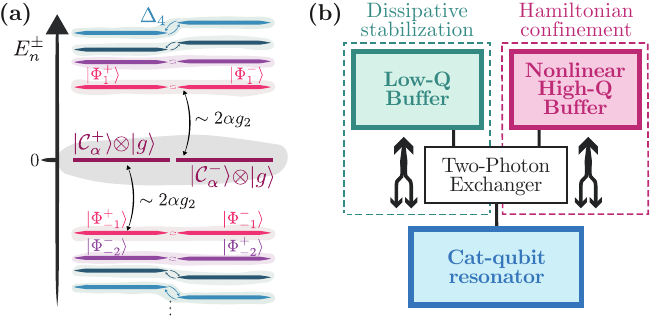}
\caption{\label{fig:tpe-intro}
(a)~Energy spectrum of the TPE Hamiltonian separated in two branches of even (left) and odd (right) parity eigenstates. The zero-energy eigenspace is the cat qubit manifold, gapped from the rest of the spectrum by an energy $\lvert E_1^\pm - E_0 \rvert \approx 2 |\alpha| g_2$. The spectrum is symmetric with respect to the zero-energy point. (b)~Schematic of a combined TPE and two-photon dissipation confinement scheme. The cat qubit resonator (blue) is coupled to both a low-Q buffer mode (green) and an anharmonic high-Q buffer mode (magenta) through an element that exchanges pairs of photons of the cat qubit mode with single photons from either buffer mode.
}
 \end{figure}

Instead of coupling the cat qubit resonator to a low-Q buffer mode, it is interesting to consider what would happen using a high-Q and highly anharmonic buffer mode. Considering only the two lowest energy levels of the buffer mode, the Two-Photon Exchange (TPE) Hamiltonian resulting from such a setup reads
\begin{equation}
    \H_\text{TPE} = (\a^2 - \alpha^2) \Hat{\sigma}_+ + \hc
\end{equation}
where $\Hat{\sigma}_\pm$ are the lowering and raising operators of the two-level buffer system. The TPE Hamiltonian features an exactly degenerate zero-energy eigenspace given by the cat qubit codespace with the buffer in its ground state, $\text{span}\{\ket{\Phi_0^+}=\ket{\mathcal{C}_\alpha^+} \ket{g}, \ket{\Phi_0^-}=\ket{\mathcal{C}_\alpha^-} \ket{g}\}$. It also preserves the photon number parity in the cat qubit resonator, such that its eigen-spectrum can be separated in two branches of even and odd parities. Solving the eigenvalue equation $\H_\text{TPE} \ket{\Phi_n^\pm} = E_n^\pm \ket{\Phi_n^\pm}$, the other eigenstates of the TPE Hamiltonian are found to be
\begin{equation}\label{eq:tpe-eigenstates}
    \begin{split}
        \ket{\Phi_n^\pm} =&~\frac{1}{\sqrt{2}} \left(\ket{\phi_n^\pm} \ket{g} + \ket{\widetilde{\phi}_n^\pm} \ket{e}\right), \text{~} E_n^\pm = \sqrt{e_n^\pm}\\
        \ket{\Phi_{-n}^\pm} =&~\frac{1}{\sqrt{2}} \left(\ket{\phi_n^\pm} \ket{g} - \ket{\widetilde{\phi}_n^\pm} \ket{e}\right), \text{~} E_{-n}^\pm = -\sqrt{e_n^\pm}
    \end{split}
\end{equation}
where $\{\ket{\phi_n^\pm}\}_{n=0}^\infty$ are excited eigenstates of the Kerr Hamiltonian $\H_\text{Kerr}=(\adagd - \alpha^{* 2})(\a^2-\alpha^2)$ associated to eigenvalues $e_n^\pm$, thus  $\H_\text{Kerr} \ket{\phi_n^\pm} = e_n^\pm \ket{\phi_n^\pm}$. Furthermore, $\{\ket{\widetilde{\phi}_n^\pm}\}_{n=1}^\infty$ are eigenstates of the reversed Kerr Hamiltonian $\widetilde{H}_\text{Kerr} = (\a^2 - \alpha^2) (\adagd - \alpha^{* 2})$ such that $\widetilde{H}_\text{Kerr} \ket{\widetilde{\phi}_n^\pm} = e_n^\pm \ket{\widetilde{\phi}_n^\pm}$, with the same eigenvalues. Eigenstates of these two Hamiltonians are linked by the relation $\ket{\widetilde{\phi}_n^\pm} = (\a^2 - \alpha^2) \ket{\phi_n^\pm} / \sqrt{e_n^\pm}$. The spectrum of the TPE Hamiltonian, depicted on Figure~\ref{fig:tpe-intro}(a) for $|\alpha|^2 = 8$, is thus closely related to the Kerr spectrum through Eq.~\eqref{eq:tpe-eigenstates}.

First and foremost, the TPE Hamiltonian thus inherits an energy gap of approximately $2|\alpha|$ between the cat space and the rest of the spectrum. As such, it can thus be used for Hamiltonian cat confinement. In the notations of Eq.~\eqref{eq:mastereq}, the TPE confinement super-operator reads $\mathcal{L}_\text{conf} = -i[\H_\text{TPE}, \cdot]$ with confinement rate $g := g_2$. While in our analysis and simulations we consider the buffer mode to be a two-level system, in practice, it can be realized using a sufficiently anharmonic mode. As shown in Section~\ref{sec:exp}, this anharmonicity should be large enough compared to the TPE Hamiltonian gap to avoid the appearance of spurious eigenstates inside the gap.

 \begin{figure}[t!]
         \centering
         \includegraphics[width = 1.0\columnwidth]{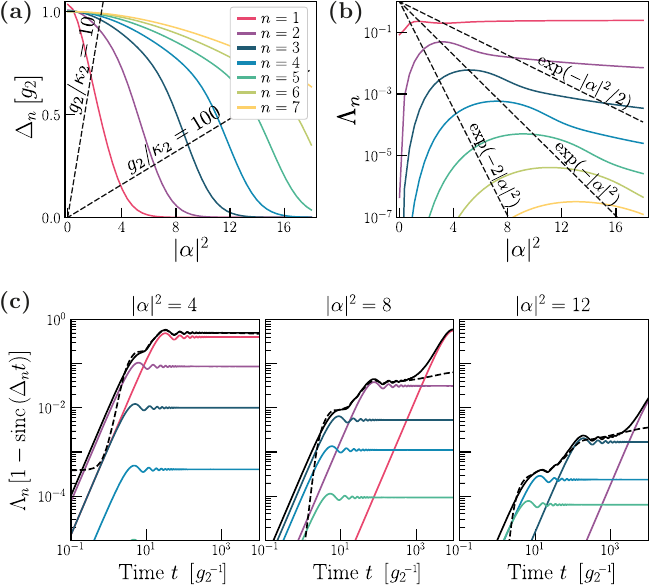}
\caption{\label{fig:tpe-degeneracies}
(a)~Energy level spacing between even and odd parity eigenstates of the TPE Hamiltonian $\Delta_n = E_n^- - E_n^+$ in units of $g_2$. Dashed lines represent the threshold $\Delta_n = \kappa_\text{conf}$ at different values of $g_2 / \kappa_2$. Only eigenstates below this threshold benefit from the additional two-photon dissipation scheme, meaning that $\Delta_n / \kappa_\text{conf} \ll 1$ in Eq.~\eqref{eq:tpe-gammabitflip-est}. (b)~Relative overlap between the $n$-th pair of excited TPE eigenstates and a cat qubit state deformed by thermal or dephasing noise-induced leakage (cf. Appendix~\ref{sec-apdx:bitflipest}). Dashed lines represent the threshold $\kappa_l \Lambda_n = \kappa_l \exp(-c |\alpha|^2)$, which one may want to impose with certain values of $c$ in order to achieve exponential suppression of bit-flip errors. All eigenstates above the threshold are susceptible to contribute to a bit-flip rate larger than the target rate via Eq.~\eqref{eq:tpe-gammabitflip-est}. (c)~Individual terms of the sum of Eq.~\eqref{eq:tpe-gammabitflip-est} with respect to time. The solid black line represents the sum of all terms. The dashed black line represents $p_X(t) / \kappa_l t$ extracted from a numerical simulation of TPE confinement with single-photon loss $\kappa_1 = 10^{-3} g_2$ and thermal noise $n_\text{th} = 10^{-2}$.
}
 \end{figure}

Second, like for the Kerr Hamiltonian, excited eigenstates of even and odd photon number parity have different energies, with each spacing closing exponentially at sufficiently large $|\alpha|$. This contributes to bit-flip errors via the same mechanism as the Kerr Hamiltonian, \ie~with an expression similar to Eq.~\eqref{eq:kerr-bitflip}. However, unlike the Kerr Hamiltonian, the associated energy gaps $\Delta_n = \sqrt{e_n^-} - \sqrt{e_n^+}$ for the TPE Hamiltonian remain uniformly bounded, as shown on Figure~\ref{fig:tpe-degeneracies}(a). This can be understood as a direct consequence of the square root: denoting $e_n^{\pm} = \bar{e}_n \pm \delta_n / 2$ and considering $\bar{e}_n \gg \delta_n$ to expand the square root, we get $\Delta_n \approx \delta_n / 2 \sqrt{\bar{e}_n}$; thus $\delta_n$ increasing more slowly than $\sqrt{\bar{e}_n}$ is sufficient to keep $\Delta_n$ bounded. Figure~\ref{fig:tpe-degeneracies}(b) shows the associated relative overlap between a cat qubit state deformed by leakage and the $n$-th pair of TPE eigenstates.

The boundedness of $\Delta_n$ has major consequences when considering the combination of Hamiltonian confinement with two-photon dissipative confinement, allowing us to fully benefit from the advantages of both schemes. Indeed, consider the bit-flip rate estimation
\begin{equation}\label{eq:tpe-gammabitflip-est}
    \begin{split}
        \Gamma_\text{bit-flip} =&~\kappa_1 |\alpha|^2 \exp\left(-4|\alpha|^2\right) \\
        &+\kappa_l \exp \left(-2|\alpha|^2 \right) \\
        &+\kappa_l \sum_{n \neq 0} \Lambda_n \left[ 1 - \frac{\sin(\Delta_n / \kappa_\text{conf})}{\Delta_n / \kappa_\text{conf}} \right]
    \end{split}
\end{equation}
which is the equivalent of Eq.~\eqref{eq:kerr-gammabitflip-est} for TPE confinement. Similarly, $\kappa_l \Lambda_n$ is the rate at which pairs of TPE eigenstates are populated by codespace leakage, where $\Lambda_n = \sum_\pm |\braket{\Phi_n^\pm | \alpha,1}|^2 / 2 = \lambda_n / 2$. Indeed, these overlaps are exactly half of their Kerr equivalents but they are twice as many due to the symmetry of the TPE energy spectrum (see Appendix~\ref{sec-apdx:bitflipest}).

Contrarily to the Kerr case, the last bracket can now be made small uniformly for all $n$ by taking $\kappa_\text{conf} \gg g_2$. This criterion recasts into $g_2 / \kappa_2 \ll 4 |\alpha|^2$. Concretely, whereas we have argued that an exponential suppression of bit-flip errors at a rate $\exp(-|\alpha|^2)$ and for $|\alpha|^2 = 12$ requires $\kappa_2\gg 0.42K$ for a Kerr confinement, the same kind of computation leads to a requirement of $\kappa_2\gg .014g_2$ in the case of TPE confinement.

\begin{figure}[t!]
         \centering
         \includegraphics[width = 1.0\columnwidth]{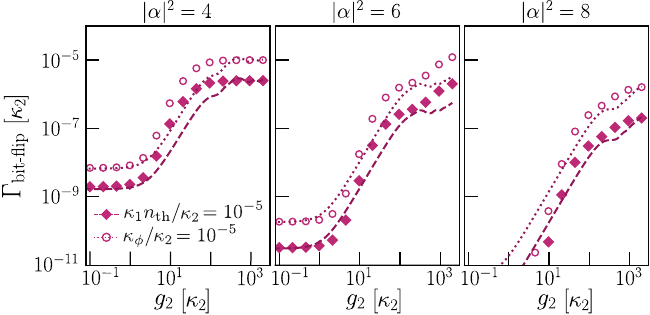}
\caption{\label{fig:tpe-diss-combined}
Bit-flip error rate of an idle cat qubit confined by a combined TPE Hamiltonian and two-photon dissipation scheme for increasing cat sizes. Both axes are graduated in units of $\kappa_2$. The cat qubit is subject to single-photon loss of amplitude $\kappa_1  = 10^{-3}\kappa_2$, with additional thermal excitations $\kappa_1 n_{\text{th}} = 10^{-5}\kappa_2$~(diamonds, dashed lines) or pure dephasing $\kappa_\phi = 10^{-5}\kappa_2$~(circles, dotted lines). The exponential bit-flip suppression is restored at low enough $g_2$, an order of magnitude earlier than for the Kerr Hamiltonian. Markers indicate numerical simulation data, lines represent the formula \eqref{eq:tpe-gammabitflip-est}.
}
 \end{figure}

These arguments are validated by numerical simulations. Figure~\ref{fig:tpe-diss-combined} shows the bit-flip rate associated to the combined TPE and two-photon dissipation confinement schemes. Similarly to the combined Kerr scheme shown in Fig.~\ref{fig:kerr-diss-combined}, the cat qubit is idling and the amplitude of the TPE confinement is varied. The same noise parameters and number of photons are used. The markers indicate numerical simulation data, the analytical predictions represented with dashed and dotted lines correspond to Eq.~\eqref{eq:tpe-gammabitflip-est}, which appear to match quite well. Qualitatively, the plots resemble Fig.~\ref{fig:kerr-diss-combined} of the Kerr Hamiltonian. For $g_2 / \kappa_2 \ll 1$, the purely dissipative cat qubit regime is retrieved with a clear exponential suppression at rate $\exp(-2 |\alpha|^2)$. In the opposite regime of $g_2 / \kappa_2 \gg 1$, the TPE cat qubit regime is retrieved. Between these two regimes, a smooth transition occurs. Quantitatively however, the transitions occur at much larger $g_2 / \kappa_2$ values, as the horizontal scale on Fig.~\ref{fig:tpe-diss-combined} is shifted to the left by more than one order of magnitude.

\begin{figure}[t!]
         \centering
         \includegraphics[width = 0.95\columnwidth]{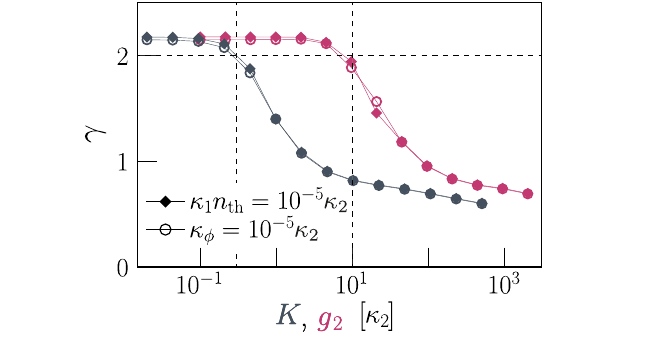}
\vspace{-0.1cm}
\caption{\label{fig:gamma-bitflip}
Exponential suppression factor of bit-flip errors $\gamma$, such that $\Gamma_\text{bit-flip} \propto \exp(-\gamma |\alpha|^2)$, for an idle cat qubit confined with a combined dissipative and Hamiltonian scheme, for either a TPE Hamiltonian~(purple) or a Kerr Hamiltonian~(grey), in units of $\kappa_2$. The $\gamma$ value is computed from an exponential fit over the range $2 \leq |\alpha|^2 \leq 12$.
The cat qubit is subject to single-photon loss of amplitude $\kappa_1 = 10^{-3}\kappa_2$, in addition with thermal excitations $\kappa_1 n_{\text{th}}  = 10^{-5}\kappa_2$~(diamonds) or pure dephasing $\kappa_\phi  = 10^{-5}\kappa_2$~(circles). Dashed lines show the approximate boundary of the $\gamma \geq 2$ exponential suppression regime, at $g_2 / \kappa_2 = 10$ and $K / \kappa_2 = 0.3$ respectively.}
 \end{figure}

Another way to look at the data is presented in Figure~\ref{fig:gamma-bitflip}. The vertical axis represents the exponential suppression factor $\gamma$ such that $\Gamma_\text{bit-flip} \propto \exp(-\gamma |\alpha|^2)$, for an exponential fit over the range $2 \leq |\alpha|^2 \leq 12$; see Appendix~\ref{sec-apdx:gamma} for details and plots about this fit. This exponential suppression factor is evaluated as a function of $g_2 / \kappa_2$ and $K / \kappa_2$ respectively for the TPE and Kerr Hamiltonian confinements. Both cases are thus simulated in combination of dissipative confinement, and with the same error channels as on the previous figures. For both combined Kerr and combined TPE confinement, a smooth transition is observed from the $\gamma \geq 2$ regime (purely dissipative confinement) to the $\gamma \rightarrow 0$ regime (bit-flip of order $\kappa_l$ induced by strong Hamiltonian contribution in Eqs.~\eqref{eq:kerr-gammabitflip-est} or \eqref{eq:tpe-gammabitflip-est}). Working points which maximize the Hamiltonian confinement without compromising the bit-flip protection can be identified just before this transition starts, respectively at $g_2 / \kappa_2 \approx 10$ and $K / \kappa_2 \approx 0.3$. As already discussed, the Kerr Hamiltonian working point is significantly lower than the TPE one due to the worse scaling of its energy level spacing.

Note that the Hamiltonian gaps of the Kerr and TPE Hamiltonian scale as $4|\alpha|^2$ and $2|\alpha|$ respectively. At the working points computed here, this implies that the TPE Hamiltonian gap is still $5$ to $10$ times larger than the Kerr one for a mean number of photon $2 \leq |\alpha|^2 \leq 12$. This should enable us to truly benefit from the combination of both confinement strategies, namely retaining the exponential bit-flip suppression induced by the dissipative stabilization, with a TPE Hamiltonian strong enough to drastically improve the speed of quantum gates. Simulations in the following sections further indicate that gate performances put $K$ and $g_2$ on an equal footing instead of the Hamiltonian gaps, thus further favoring the TPE confinement scheme.

Before ending this section, we must say a few words about the experimental implementation of the TPE confinement. The TPE Hamiltonian is also interesting in this respect. Its similarity to the two-photon dissipation indicates  that they can be engineered simultaneously and through simple modifications of the dissipative setup~\cite{lescanne2020exponential}. Figure~\ref{fig:tpe-intro}(b) shows a schematic of a cat qubit with both types of confinement schemes. The resonator that encodes the cat qubit (blue) is coupled to a low-Q buffer (green) and to a nonlinear high-Q buffer (magenta) through a two photon exchanger. A single element can be used to engineer the two TPE couplings, with the low-Q and high-Q buffers, thus inducing a combined dissipative-conservative confinement which has the benefit of protecting against leakage from the code space due to thermal and dephasing noise, but also against spurious Hamiltonians with energy scales small compared to the TPE energy gap. 

In the following sections, bias-preserving single-qubit Z gates and two-qubit CNOT gates with combined confinement schemes are investigated at the working points identified on Figure \ref{fig:gamma-bitflip}. The Hamiltonian confinements will then show their benefits in rejecting the non-ideal terms present in the gate Hamiltonians, while the dissipative confinement maintains protection against perturbation channels like $\kappa_1$ and $\kappa_\phi$. Note that the addition of a buffer mode into the system leads to additional noise sources for TPE Hamiltonian confinement. The effect of these sources is thoroughly investigated in Appendix~\ref{sec-apdx:circuit-derivation}, confirming that our conclusions should still hold under realistic conditions.

\section{Z gate with combined Confinement Schemes}\label{sec:zgate}

In quantum computing architectures with biased-noise qubits, it is essential that the error bias is conserved throughout the operation of the device. A quantum gate with cat qubits should therefore preserve the exponential suppression of bit-flip errors. In this paper, we focus on the bias-preserving realization of single-qubit Z gates and two-qubit CNOT gates~\cite{guillaud2019repetition,puri2020bias}. Similarly to~\cite{guillaud2019repetition}, the proposed CNOT gate can be extended to a three qubit Toffoli (Toffoli) gate which together with appropriate state preparation and measurements provide the building blocks of a hardware-efficient fault-tolerant universal quantum computer based on repetition cat qubits~\cite{guillaud2019repetition,chamberland2020building}. Single-qubit X gates of angle $\pi$ are also required for universality but are trivial to implement either in software, by commuting the Pauli gate with the circuit, or in hardware, with a half-period delay on the oscillator.

In this section, we focus on the case of single-qubit Z gate with a combined dissipative and Hamiltonian confinement. Section \ref{sec:cnot} introduces a new design for the CNOT gate under such combined confinement. The addition of a Kerr or TPE confinement compared to fully dissipative cat qubits offers the perspective of improved gate speed, while the two-photon dissipation stabilizes the code manifold and protects it from leakage-induced bit-flip errors.

Similarly to~\cite{mirrahimi2014dynamically,puri2017engineering} and the experimental realizations of~\cite{touzard2018coherent,grimm2020stabilization}, a rotation around the Z axis of the  cat qubit can be performed  via the application  of a resonant drive on the cat qubit resonator. Indeed, taken alone this would displace the cat in phase space, but adding it to a confinement process essentially induces a phase accumulation between the different regions of phase space where the state remains confined, thus essentially between $\ket{\alpha} \approx \ket{0_L}$ and $\ket{-\alpha} \approx \ket{1_L}$. In an appropriate rotating frame, the system is therefore subject to the following master equation during the gate:
\begin{equation}\label{eq:zgate-mastereq}
    \frac{\dd \hrho}{\dd t} = -i \left[\varepsilon_Z(t) \adag + \hc, \hrho \right] -i g\left[\H_\text{conf}, \hrho \right] + \kappa_2 \mathcal{D}[\LD]\hrho \; ,
\end{equation}
where $g \H_\text{conf} \equiv - K \H_\text{Kerr}$ or $g \H_\text{conf} \equiv g_2 \H_\text{TPE}$ for combined Kerr and combined TPE confinements respectively. Furthermore $\varepsilon_Z(t)$ denotes the complex amplitude of the resonant drive that can be slowly varying in time, turning on and off the logical $Z$ rotation. 

A fully dissipative gate is retrieved at $g=0$. This situation can be analyzed with Zeno dynamics at various orders of the small parameter $|\varepsilon_Z/\kappa_2|$. The state always remains  $|\varepsilon_Z/\kappa_2|$-close to the cat-qubit subspace and, as argued in~\cite{guillaud2019repetition}, the bit-flip errors remain exponentially suppressed during the gate. At first order, the $\varepsilon_Z$ drive induces an effective Zeno dynamics that rotates the state of the qubit around its Z axis of the Bloch sphere at a speed given by $4 \Re(\alpha^* \varepsilon_Z(t))$~\cite{mirrahimi2014dynamically}, thus performing the gate. At the second order, it leads to effective phase decoherence $\mathcal{D}[Z]$, in other words phase-flip errors. These so-called non-adiabatic errors thus scale as $T_\text{gate} \varepsilon_Z^2 / \kappa_2 \propto 1/(T_\text{gate}\kappa_2)$ since $\varepsilon_Z \propto 1/T_\text{gate}$ where $T_\text{gate}$ is the gate time. However, typical decoherence channels on the cat qubit resonator, like the dissipation channels with $\kappa_1$ and $n_{\text{th}}$ described in the simulations of Sections \ref{sec:bitflip-confinement} and \ref{sec:tpe}, induce direct phase errors, which scale linearly with the gate time as $T_\text{gate}$. The trade-off between these two effects yields an optimal gate time at which gate fidelity is maximal~\cite{guillaud2019repetition, chamberland2020building}.

A fully Hamiltonian gate is instead retrieved at $\kappa_2 = 0$. This situation can be analyzed from the viewpoint of an adiabatically varying Hamiltonian. Taking advantage of the Hamiltonian gap, the adiabatic theorem ensures that transitions outside the cat qubit subspace can be suppressed exponentially in the gate time \cite{nenciu1993linear}, if the Hamiltonian varies smoothly enough. Since the evolution is purely Hamiltonian, this implies that gate-induced errors can be suppressed exponentially in $T_\text{gate}$. The prefactors of these errors can be further improved drastically by using superadiabatic pulse designs as proposed in Ref.~\cite{xu2021engineering}. It thus seems that gates can be performed orders of magnitudes faster with Hamiltonian confinement than with a fully dissipative confinement scheme, for which non-adiabatic errors scale linearly in $1/T_\text{gate}$. However, as we have shown in the previous sections, a fully Hamiltonian confinement does not preserve the error bias of cat qubits under typical dissipation channels, and is therefore irrelevant on its own.

With a combined dissipative and Hamiltonian confinement, the phase-flip performance of Z gates can be improved without compromising bit-flip errors. From a general viewpoint, the phase flip error probability of a Z gate can be modeled as
\begin{equation}\label{eq:phase-error-model}
    p_Z = \kappa_1 |\alpha|^2 T_\text{gate} + p_Z^\text{NA}
\end{equation}
where the first term is identical to the idling qubit and represents errors due to the dominant dissipation channel, namely single-photon loss in the resonator mode, while $p_Z^\text{NA}$ corresponds to ``non-adiabatic'' phase errors induced by the gate operation. For a Z gate with combined Kerr and two-photon dissipation, the gate-induced phase-flip error probability is derived in Appendix~\ref{sec-apdx:nonadiab-errors}, assuming a constant Hamiltonian drive throughout the gate and neglecting fast transients when switching it on and off. The gate-induced phase-flip error is then given by
\begin{equation}\label{eq:pz-na-kerr}
    p_Z^\text{NA} = \frac{1}{1 + \frac{4K^2}{\kappa_2^2}} \frac{\theta^2}{16|\alpha|^4 \kappa_2 T_\text{gate}}
\end{equation}
where $\theta$ is the angle of the Z gate. This expression contains the dominant error, which is closer to the fully dissipative case than to the purely Hamiltonian case, with a linear scaling in $1/T_\text{gate}$. Compared to the model of Ref.~\cite{chamberland2020building} for fully dissipative gates, this equation features an additional prefactor that is tunable through the ratio between Kerr non-linearity and dissipative confinement rate. Consequently, non-adiabatic phase errors are suppressed quadratically with the addition of a Kerr confinement Hamiltonian compared to the fully dissipative design. One should bear in mind though that this additional confinement, in presence of noise mechanisms such as thermal excitation and photon dephasing, induces additional bit-flip errors as discussed in previous sections, which are kept in check up to the working point provided in Fig.~\ref{fig:gamma-bitflip}. For larger Hamiltonian confinement rates than this working point, phase errors can be further reduced at the cost of an increase in bit-flip errors. In this case, an optimization could be performed at the level of the logical quantum error correcting code to find an optimal working point.

For the combined TPE and two-photon dissipation confinement, the gate-induced phase-flip error probabilities are numerically fitted with
\begin{equation}\label{eq:pz-na-tpe}
    p_Z^\text{NA} = \frac{1}{1 + \frac{4g_2^2}{\kappa_2^2}} \frac{\theta^2}{16|\alpha|^4 \kappa_2 T_\text{gate}} + \frac{\theta^2}{32 |\alpha|^4 g_2^2 T_\text{gate}^2} \; .
\end{equation}
The first term is similar to the combined Kerr gate, with a quadratic suppression of non-adiabatic phase errors in the ratio between Hamiltonian and dissipative confinement rates. The second term appears to be specific to the TPE confinement and decreases quadratically in $g_2 T_\text{gate}$ independently of $\kappa_2$, as long as $g_2 \gtrsim \kappa_2$.

\begin{figure}[t!]
         \centering
         \includegraphics[width = 1.0\columnwidth]{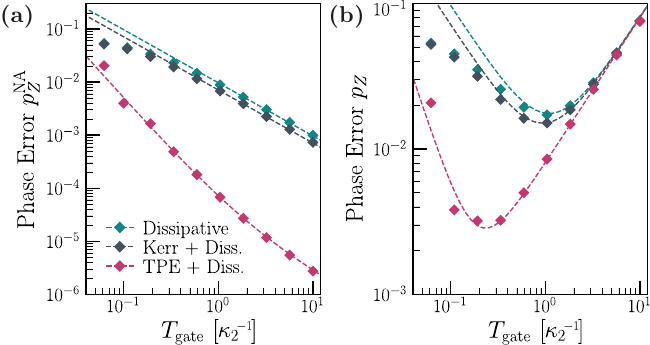}
\caption{\label{fig:zgate-combined}
    (a)~Gate-induced phase errors for a Z gate at $|\alpha|^2 = 8$ with different confinement schemes: two-photon dissipation (green), combined Kerr and two-photon dissipation at $K / \kappa_2 = 0.3$ (grey), and combined TPE and two-photon dissipation at $g_2 / \kappa_2 = 10$ (magenta). (b)~Total phase errors for the same Z gates, assuming single-photon loss at rate $\kappa_1  = 10^{-3}\kappa_2$, thermal noise $n_\text{th} = 10^{-2}$ and pure dephasing $\kappa_\phi = 10^{-5}\kappa_2$. The combination of TPE and dissipative confinement allows us to win almost one order of magnitude on the phase error, and operates optimally almost one order of magnitude faster. For both plots, a constant Hamiltonian drive $\varepsilon_Z$ is used and gate times are in units of $1/\kappa_2$. Markers indicate numerical data, dashed lines indicate analytical fits.
}
\end{figure}


In Figure~\ref{fig:zgate-combined}(a), these gate-induced phase errors are represented for three  confinement schemes: a two-photon dissipation scheme (in green), a combined Kerr and two-photon dissipation scheme at the working point $K / \kappa_2 = 0.3$ (in grey) and a combined TPE and two-photon dissipation scheme at the working point $g_2 / \kappa_2 = 10$ (in magenta). We remind that, as shown in Fig.~\ref{fig:gamma-bitflip}, these working points correspond to maximal Hamiltonian confinement strengths at which the bit-flip errors remain suppressed with a rate of order $\exp(-2|\alpha|^2)$ for $2 \leq |\alpha|^2 \leq 12$. The mean number of photons is fixed at $|\alpha|^2 = 8$ and the gate time varies along the horizontal axis. Markers show numerical data, while lines show analytical fits as given by Eqs.~\eqref{eq:pz-na-kerr} and \eqref{eq:pz-na-tpe}. 

For the gate with a fully dissipative confinement, we find similar results as in the literature~\cite{guillaud2019repetition, chamberland2020building, xu2021engineering} with non-adiabatic errors suppressed linearly in the gate time. For the combined Kerr and dissipative confinement, almost no gain in performance is found compared to the dissipative one because of the low Kerr working point at $K / \kappa_2 = 0.3$. The model of Eq.~\eqref{eq:pz-na-kerr} predicts a reduction by a factor of about $1.36$ on non-adiabatic errors. On the other hand, the combined TPE and dissipative confinement shows a significant gain in performance with a factor of about 400 compared to the fully dissipative confinement, in the limit of large gate times.

Figure~\ref{fig:zgate-combined}(b) shows the complete phase-flip error probability in the presence of the other noise sources on the cat qubit resonator, and in particular with single-photon losses at rate $\kappa_1  = 10^{-3}\kappa_2$. In this case, the optimal phase-flip error probability for Z gate drops from $p_Z^* \approx 2\%$ in the fully dissipative scheme to $p_Z^* \approx 0.3\%$ in the combined TPE and dissipative scheme. The optimal gate time $T_\text{gate}^*$ that minimizes phase errors is also much smaller with the combined TPE scheme than with the fully dissipative one, hence leading to faster gate designs. For lower values of $\kappa_1 / \kappa_2$, the performance gain compared to fully dissipative gates is expected to increase, up to the factor of $400$ characterizing the non-adiabaticity error.

\begin{figure}[t!]
         \centering
         \includegraphics[width = 1.0\columnwidth]{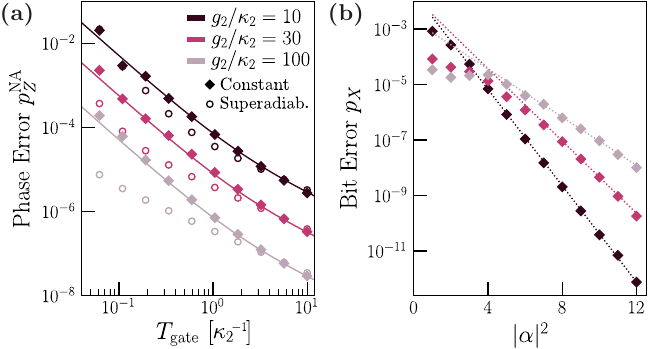}
\caption{\label{fig:zgate-tpedis}
    (a)~Gate-induced phase errors for a Z gate at $|\alpha|^2 = 8$ for a combined TPE and two-photon dissipation confinement, at varying confinement ratios of $g_2 / \kappa_2$. Diamonds indicate a constant gate Hamiltonian drive, while circles indicate a gaussian-type gate Hamiltonian drive with first order superadiabatic correction, as in Ref.~\cite{xu2021engineering}. Lines indicate the analytical fit Eq.~\eqref{eq:pz-na-tpe} in the constant drive regime. Gate times are in units of $1/\kappa_2$. (b)~Total bit-flip errors for a Z gate under the same confinement schemes, with single-photon loss at rate $\kappa_1 = 10^{-3}\kappa_2$, thermal noise $n_\text{th} = 10^{-2}$ and pure dephasing $\kappa_\phi = 10^{-5}\kappa_2$. Markers indicate numerical data, and dotted lines indicate exponential fits for $2 \leq |\alpha|^2 \leq 12$; the corresponding exponential suppression factors are the values $\gamma$ represented on Fig.~\ref{fig:gamma-bitflip}.
}
\end{figure}


In Figure~\ref{fig:zgate-tpedis}, we further explore different working points for the combined TPE and dissipative confinement, at increasing $g_2 / \kappa_2$ values. Plot (a) shows non-adiabatic phase flip errors while plot (b) shows the corresponding bit-flip errors, with same noise sources as in the Fig.~\ref{fig:zgate-combined}(b). As expected, non-adiabatic phase flip errors are suppressed quadratically with $g_2 / \kappa_2$, while the exponential suppression of bit-flip errors is continuously degraded from $\gamma \geq 2$ as $g_2 / \kappa_2$ increases above the working point. The dotted lines on plot (b) follow the values of $\gamma$ shown in Figure~\ref{fig:gamma-bitflip}. To find the optimal error rate on the logical level, an optimization on the level of the quantum error correcting code should be performed. 

Figure~\ref{fig:zgate-tpedis}(a) also shows Z gate simulations with a superadiabatic pulse design (carved out circles). The applied drive amplitude is an adaptation of Ref.~\cite{xu2021engineering} to the TPE Hamiltonian confinement and reads $\varepsilon_Z(t) \propto \Omega_G(t) + \ddot{\Omega}_G(t) / E_1^2$ where $\Omega_G$ is a second-order gaussian pulse and $E_1$ is the TPE Hamiltonian gap. The simulations show that such drive designs can further improve the performance of Z gates, but their benefits especially show up beyond the working point $g_2 / \kappa_2 = 10$. In other words, the presence of the necessary dissipative confinement in order to counter bit flips induced by $\kappa_1,\; n_\text{th}$ and $\kappa_\phi$ processes, appears to limit the benefits of superadiabatic drives in this setting. Whether the compatibility of combined confinement with superadiabatic drives can be improved, remains a question for future research.

\section{CNOT Gate with Combined Confinement Schemes}\label{sec:cnot}

The two-qubit CNOT gate is one of the most elementary entangling gates and holds a major role in the operation of a quantum computer. In error correcting codes, it is used to map the error syndromes on ancilla qubits that are subsequently measured. The success of an error correction process is therefore mainly limited by the infidelity of CNOT gates. Indeed, the effective logical error probability is more sensitive to physical CNOT fidelity than to any other component of the error correcting code, such as ancilla preparation or measurement~\cite{fowler2012surface}. It is therefore crucial to engineer CNOT gates that can reach fidelities well above the error correction threshold. In this section, we first review the proposals on the design of CNOT gates for confined cat qubits. We then provide a new proposal that can achieve fast gate speeds, with drastically improved phase fidelity compared to dissipative schemes, and without compromising the bit-flip protection like fully Hamiltonian schemes.

For dissipative cat qubits~\cite{guillaud2019repetition, guillaud2021error,chamberland2020building}, CNOT gates can be engineered through a slow variation of the dissipative confinement parameters, such that the target cat-qubit manifold experiences a slow rotation in the harmonic oscillator phase space conditionally on the state of the control cat-qubit. Once the conditional rotation reaches an angle $\pi$, the rotated and original cat-qubit manifolds are again superimposed, but with a logical X-rotation between the states that they encode. More concretely, this process is effectively described by the time-dependent Lindblad operators
\begin{subequations}\label{eq:diss-cnot-ops}
    \begin{equation}
        \LD^{(\text{co})} = \a_\text{co}^2 - \alpha^2  \; ,
    \end{equation}\vspace{-1em} 
    \begin{equation}
        \LD^{(\text{ta})} (t) = \a^2_\text{ta} - \alpha^2 + \frac{\alpha}{2} (\e^{2i\varphi(t)} - 1) (\a_\text{co} - \alpha)
    \end{equation}
\end{subequations}
where $\a_\text{co/ta}$ are the annihilation operators of the control and target cat qubit resonators, $\varphi(t) = \pi t / T_\text{gate}$ is the rotation angle of the target cat-qubit codespace in the harmonic oscillator phase space, and $T_\text{gate}$ is the gate time. Indeed, at fixed $t$, the steady state of these dissipators is a four-dimensional subspace spanned by the coherent states $\ket{\alpha}_\text{co} \otimes \ket{\pm \alpha}_\text{ta}$ and $\ket{-\alpha}_\text{co} \otimes \ket{\pm \alpha \e^{i \varphi(t)}}_\text{ta}$. Assuming that the state of the system follows exactly the steady state subspace throughout the gate, a state $\ket{\alpha}_\text{co} \otimes \ket{\pm \alpha}_\text{ta} \approx \ket{0_L}_\text{co} \otimes \ket{0/1_L}_\text{ta}$ would thus not move, while a state $\ket{-\alpha}_\text{co} \otimes \ket{\pm \alpha}_\text{ta} \approx \ket{1_L}_\text{co} \otimes \ket{0/1_L}_\text{ta}$ would have moved after time $T_\text{gate}$ to $\ket{-\alpha}_\text{co} \otimes \ket{\mp \alpha}_\text{ta} \approx \ket{1_L}_\text{co} \otimes \ket{1/0_L}_\text{ta}$, effectively achieving a CNOT gate. However, such an assumption is only fulfilled in the limit of infinitely slow gates. In realistic conditions, the state of the harmonic oscillator lags behind the stabilized manifold, inducing errors.

To effectively reduce these errors, a so-called feedforward Hamiltonian can be added throughout the gate. This approach, which can be seen as a shortcut to adiabaticity for open quantum systems~\cite{alipour2020shortcuts}, drives the oscillator state in such a way as to maintain it within the time-dependent dissipators steady state manifold, hence suppressing non-adiabatic errors. In its ideal form, a feedforward Hamiltonian for the cat-qubit CNOT gate would read
\begin{equation}\label{eq:cnot-feedforward-ideal}
    \H_{\text{CX}, \text{ideal}} = \dot{\varphi} \dm{1_L}{1_L}_\text{co} \otimes \adag_\text{ta} \a_\text{ta} + \H_{\text{s}} \; ,
\end{equation}
where the first term makes the target qubit rotate conditionally on the control being in the $\ket{1_L}_\text{co}$ logical state, and $\H_{\text{s}}$ with $\H_{\text{s}}\ket{1_L}_\text{co} = \H_{\text{s}} \ket{0_L}_\text{co} = 0$  allows an arbitrary Hamiltonian when the control qubit is outside the code space~\cite{guillaud2019repetition,chamberland2020building}. To approach such ideal feedforward Hamiltonian with realistic Hamiltonians, an approximate version is proposed~\cite{guillaud2019repetition, puri2020bias}
\begin{equation}\label{eq:cnot-feedforward}
    \H_{\text{CX}} = -\varepsilon_{\text{CX}} (\a_\text{co} + \adag_\text{co} - 2\Re(\alpha)) (\adag_\text{ta} \a_\text{ta} - |\alpha|^2).
\end{equation}
where $\varepsilon_{\text{CX}} = \dot{\varphi} / 4\Re(\alpha)$. Indeed, at least for large $\alpha$, the left-hand side bracket approximately evaluates to $0$ if the control qubit is in the $\ket{\alpha}_\text{co}$ state, and to $-4\Re(\alpha)$ if the control qubit is in the $\ket{-\alpha}_\text{co}$ state (to see this, consider \eg~the mean value of the quadrature operator $\a_\text{co}+\adag_\text{co}$). This Hamiltonian can further be engineered through a four-wave mixing element coupled to cat qubit resonators as demonstrated in~\cite{touzard2019gated}. The similarity of the Hamiltonian acting on the control qubit with the Z gate Hamiltonian is no coincidence. Indeed, in a dual viewpoint the CNOT gate corresponds to a Z rotation of the control qubit, conditioned on the logical phase of the target qubit. Since logical phase here corresponds to photon-number parity of the cat-qubit, this dual viewpoint is clearly visible in $\H_{\text{CX}}$.

For Kerr cat qubits~\cite{puri2020bias}, the CNOT gate can be realized through the slowly varying Hamiltonian 
\begin{equation}\label{eq:cnot-kerr-hamil}
    \H = -K \left(\LD^{(\text{co})}\right)^\dagger \LD^{(\text{co})} -K \left(\LD^{(\text{ta})}(t)\right)^\dagger \LD^{(\text{ta})}(t) + \H_{\text{CX}} \; .
\end{equation}
The first two terms feature a time-dependent four-dimensional subspace of ground states following the same evolution as for the dissipatively confined gate, and the third term is the feedforward Hamiltonian designed to drive the oscillator state along with this time-dependent steady state manifold. In this case, the adiabaticity condition ensuring accurate gate operation reads $\dot{\varphi} \ll 4 |\alpha|^2 K$ where the right-hand side is the Kerr Hamiltonian gap~\cite{puri2020bias}. For a fully Hamiltonian-based gate with sufficiently smooth variation of $\varepsilon_{CX}$, this should ensure exponential suppression of gate-induced errors as a function of $T_{\text{gate}}$, like for the Z gate. Its scaling can be further improved by using superadiabatic pulse designs as proposed in~\cite{xu2021engineering}, adding a feedforward  Hamiltonian term $\H_{\text{CX}}^{\text{(sa)}} \propto \dot{\varepsilon}_\text{CX} \, i (\a_c - \adag_c) (\adag_t \a_t - |\alpha|^2)$. The exponential suppression of non-adiabatic errors in the gate time is a drastic improvement from the linear scalings observed both for purely dissipative gates, and for Kerr-based CNOT gates with discontinuous on/off operation of $\varepsilon_{CX}$. Fully Hamiltonian gates are however limited by the same effects as discussed in previous sections, \ie~the important rate of bit-flip errors induced by thermal and dephasing noise.

Like for the Z gate, engineering CNOT gates with a combined Hamiltonian and two-photon dissipation confinement thus arises as a promising approach to combine the best of both worlds. However, at first glance, such a scheme appears experimentally very challenging due to the large number of Hamiltonians and dissipators that need to be engineered. For a combined Kerr and two-photon dissipation CNOT gate for instance, control and target qubit should be coupled through various mixing terms that engineer the target qubit Kerr confinement as a function of the control qubit, and also to a common buffer mode to engineer the target qubit two-photon dissipation term as depending on the control qubit. With the additional feedforward Hamiltonian, Kerr nonlinearities on each mode, and various single and two-photon drives that appear in Hamiltonian~\eqref{eq:cnot-kerr-hamil}, this leads to a daunting experimental task. 

To avoid the above complexities in engineering various Hamiltonians and dissipators, we consider a much simpler scheme where the confinement is turned off on the target cat qubit during the operation of the CNOT gate~\cite{puri2019stabilized}. This is motivated by the fact that the feedforward Hamiltonian approximately preserves the coherent states composing the target cat qubit, by making them rotate in phase space by an angle that depends on the rather well-defined quadrature value of the control qubit, and it also rigorously preserves the phase of the target qubit (as it preserves photon number). During a fast gate, the main gate-induced imperfections thus take place on the control qubit, where the feedforward drive takes a similar form as for the Z gate. Leaving the target qubit unprotected only for the short gate time is therefore not too detrimental. Indeed, local leakage will be corrected as soon as the gate ends and confinement is turned back on, well before significant bit-flips can be induced. Regarding performance, it is not even clear how much could be gained by leaving on a coupled combined confinement scheme. In absence of an efficient design for TPE-type Hamiltonian confinement, with the typical parameters of the present work, gate performances are in fact improved when switching off $\kappa_2$ on the target qubit.

 \begin{figure}[t!]
         \centering
         \includegraphics[width = 1.0\columnwidth]{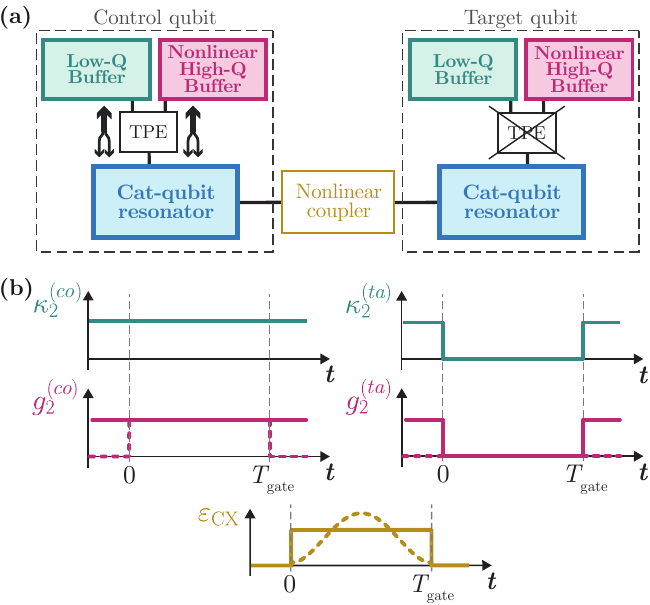}
\caption{\label{fig:cnot-schematic}
(a)~Circuit schematic of a cat qubit CNOT gate using a combined two-photon dissipation and TPE Hamiltonian confinement scheme on both qubits. The control cat qubit resonator is connected to the target cat qubit resonator through a nonlinear coupler that engineers the feedforward Hamiltonian of~\eqref{eq:cnot-feedforward}. During the CNOT gate, both Hamiltonian and dissipative confinements are turned off on the target qubit by turning off the pump of the TPE element. (b)~Amplitudes of the different dissipation and Hamiltonian terms during the CNOT gate. A constant feedforward Hamiltonian amplitude with first-order correction is assumed in full lines. Vertical dashed lines identify gate start and end times, while dashed curves show alternative approaches. More precisely, while the TPE confinement of the control cat qubit is essential during the operation of the gate to ensure a gate acceleration, it can be turned off during the idling time. Keeping it on can however be useful for compensating spurious Hamiltonian perturbations. Concerning the feedforward Hamiltonian, while simulations are performed with a constant amplitude, it is possible to benefit from superadiabatic pulse designs to further accelerate gates, similarly as in~\cite{xu2021engineering}.
}
 \end{figure}

For a combined TPE and dissipative confinement scheme, the master equation describing the evolution of the system during the CNOT gate with our proposal is thus given by
\begin{equation}\label{eq:cnot-mastereq}
    \frac{\dd \hrho}{\dd t} = -i \left[\H_{\text{CX}}, \hrho \right] -i \left[\H_\text{TPE}^{(\text{co})}, \hrho \right] + \kappa_2 \mathcal{D}[\LD^{(\text{co})}]\hrho
\end{equation}
where $\H_\text{TPE}^{(\text{co})} = g_2 (\a_\text{co}^2 - \alpha^2) \Hat{\sigma}_+ + \hc$. A schematic of such a CNOT implementation is shown in Figure~\ref{fig:cnot-schematic}. This proposal should greatly simplify the experimental design of CNOT gates as the only coupling term that remains between control and target qubits is the feedforward Hamiltonian, experimentally demonstrated in~\cite{touzard2019gated}. We next quantify the associated bit-flip and phase-flip errors.

 \begin{figure}[t!]
         \centering
         \includegraphics[width = 1.0\columnwidth]{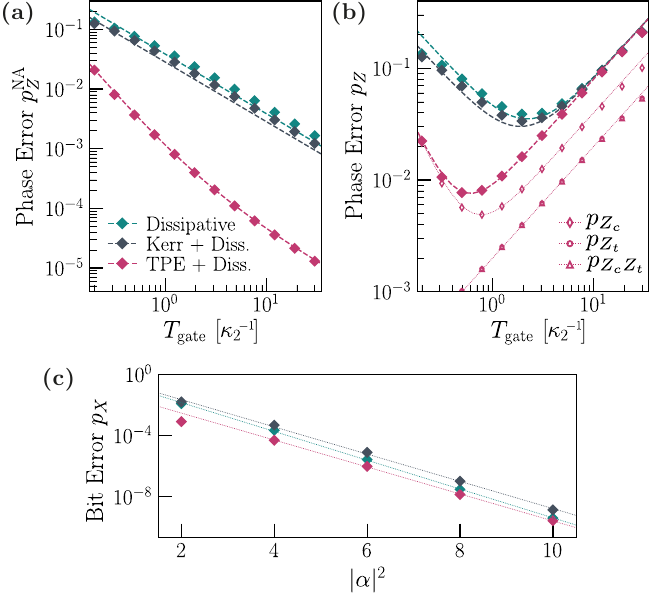}
\caption{\label{fig:cnot-combined}
(a)~Non-adiabatic phase errors of a CNOT gate at $|\alpha|^2 = 4$, with different confinement schemes: two-photon dissipation (green), combined Kerr and two-photon dissipation at $K / \kappa_2 = 0.3$ (grey), and combined TPE and two-photon dissipation at $g_2 / \kappa_2 = 10$ (magenta). For this last scheme, the confinement is turned off on the target cat qubit during the gate. (b)~Phase errors of a CNOT gate for the same confinement schemes, with single-photon loss at rate $\kappa_1 = 10^{-3}\kappa_2$, thermal noise $n_\text{th} = 10^{-2}$ and pure dephasing $\kappa_\phi = 10^{-5}\kappa_2$ on both control and target qubits. Full markers show total phase error, empty markers show the breakdown of phase errors for the combined TPE scheme. Gate times are in units of $1/\kappa_2$. (c)~Total bit-flip errors of a CNOT gate for the same confinement schemes, with noise rates as in plot (b), and at gate time $T_\text{gate} = 1/\kappa_2$. For all plots, a constant gate Hamiltonian drive is used, and markers indicate numerical data. Dashed lines indicate analytical fits, and solid lines indicate numerical fits of exponential suppression.
}
 \end{figure}

In Figure~\ref{fig:cnot-combined}, we plot the performance of CNOT gates with three different confinement schemes. The first one is the fully dissipative scheme with both a control dissipator and a time-dependent target dissipator as proposed in Ref.~\cite{guillaud2019repetition} (green), the second one is the combined Kerr Hamiltonian and two-photon dissipation acting on both control and target qubits as proposed in Ref.~\cite{puri2020bias} but at the working point $K / \kappa_2 = 0.3$ (grey), and the third one is the combined TPE and two-photon dissipation scheme as described by Eq.~\eqref{eq:cnot-mastereq} at the working point $g_2 / \kappa_2 = 10$ (magenta).

Similarly to the Z gate, Figure \ref{fig:cnot-combined}(a) first compares the `non-adiabatic' phase errors, induced by the gate operation itself in absence of any perturbations on the two qubits. As explained above, the gate only affects the phase of the control qubit. We recall that the operating points have been selected in Section \ref{sec:tpe} to anticipate reasonable protection against typical error sources. Like for the Z gate, the combined TPE scheme clearly outperforms the other two, up to a factor about $100$. For the three confinement schemes, non-adiabatic errors are fitted by the following formulas:
\noindent\begin{minipage}{\columnwidth}
    \vspace{1em}
    \begin{center}
        (Two-Photon Dissipation Confinement)
    \end{center}\vspace{-1.5em}
    \begin{equation}\label{eq:cnot-na-diss}
        p_Z^\text{NA} = \frac{\pi^2}{64 |\alpha|^2 \kappa_2 T_\text{gate}}
    \end{equation}
    \vspace{\parskip}
\end{minipage}
\noindent\begin{minipage}{\columnwidth}
    \begin{center}
        (Combined Kerr and Two-Photon Dissipation Confinement)
    \end{center}\vspace{-1.5em}
    \begin{equation}\label{eq:cnot-na-kerdiss}
        p_Z^\text{NA} = \frac{1}{1+\frac{4K^2}{\kappa_2^2}} \frac{\pi^2}{64 |\alpha|^2 \kappa_2 T_\text{gate}}
    \end{equation}
    \vspace{\parskip}
\end{minipage}
\noindent\begin{minipage}{\columnwidth}
    \begin{center}
        (Combined TPE and Two-Photon Dissipation Confinement)
    \end{center}\vspace{-1.5em}
    \begin{equation}\label{eq:cnot-na-tpediss}
        p_Z^\text{NA} = \frac{1}{1+\frac{4g_2^2}{\kappa_2^2}} \frac{\pi^2}{16 |\alpha|^2 \kappa_2 T_\text{gate}} + \frac{\pi^2}{32 |\alpha|^2 g_2^2 T_\text{gate}^2}.
    \end{equation}
    \vspace{\parskip}
\end{minipage}
The first two formulas can be derived following the shifted Fock basis approach introduced in~\cite{chamberland2020building}, similarly to the case of Z gates (Appendix~\ref{sec-apdx:nonadiab-errors}), while the third one is numerically fitted and valid as long as $g_2 \gtrsim \kappa_2$. Its analytical derivation is left for future work. All three equations scale as $|\alpha|^{-2}$ while their Z gate equivalents scale as $|\alpha|^{-4}$. This comes from the fact that the feedforward Hamiltonian $(\hat{a}_{\text{co}}+\hat{a}_{\text{co}}^\dagger)$ acting on the control qubit like for the Z gate, is here multiplied by an amplitude that scales with the variance of the photon number in the target qubit state. It can be explicitly computed in the shifted Fock basis as explained in~\cite{chamberland2020building}. Furthermore, the constant prefactor in the first term of Eq.~\eqref{eq:cnot-na-tpediss} involves $1/16$ instead of $1/64$ as in Eq.~\eqref{eq:cnot-na-diss}, Eq.~\eqref{eq:cnot-na-kerdiss} and Ref.~\cite{chamberland2020building}. This is an indirect effect of turning off the target two-photon dissipation, which lets $\H_{\text{CX}}$ induce more entanglement of the control qubit phase information with target qubit leakage. At first glance, keeping two-photon dissipation on the target qubit may thus seem beneficial to gain this prefactor. However, keeping this dissipation would also introduce additional channels for control qubit phase dissipation, reducing the benefits of its Hamiltonian confinement towards performing fast gates. As such, compared to the purely dissipative scheme, the quadratic prefactor in $g_2 / \kappa_2$ largely compensates the constant prefactor with the settings and working point of our proposal.

In Figure~\ref{fig:cnot-combined}(b), the total phase-flip error probability of CNOT gates with the same three confinement schemes are shown in the presence of typical noise sources. Similarly to \cite{guillaud2019repetition}, the latter induces phase-flip errors not only on the control qubit, but also on the target qubit, and correlated between control and target qubit. The figure hence shows the breakdown of those errors. Solely the control qubit undergoes non-adiabatic errors and the formulas are the same as in \cite{guillaud2019repetition} for purely dissipative confinement:
\begin{equation}
    \begin{split}
        p_{Z_c} =&~\kappa_1 |\alpha|^2 T_\text{gate} + p_Z^\text{NA} \\
        p_{Z_t} =&~p_{Z_c Z_t} = \frac{1}{2}\kappa_1 |\alpha|^2 T_\text{gate} \; ,
    \end{split}
\end{equation}
up to the expression of $p_Z^{\text{NA}}$. The optimal phase-flip error probability $p_{Z_c}$ remains independent of $\alpha$ when using \eqref{eq:cnot-na-tpediss} instead of \eqref{eq:cnot-na-diss}. For a realistic single-photon loss rate of $\kappa_1  = 10^{-3}\kappa_2$, this optimum is about $p_{Z_c}^* \approx 0.8\%$ for our scheme, to be compared with $p_{Z_c}^* \approx 5\%$ in the dissipative scheme. Like in the Z gate simulations, this optimal gate fidelity is furthermore obtained at a typical gate time about $5$ times faster for the combined TPE scheme. The working point selection for the combined Kerr and dissipative confinement, appears to make it barely better than purely dissipative confinement in terms of reducing non-adiabatic and hence total phase flip error. Once again, for lower values of $\kappa_1 / \kappa_2$, further performance gains can be expected compared to fully dissipative gates, up to the factor of about $100$ gained on the non-adiabaticity error at large gate times.

In Figure~\ref{fig:cnot-combined}(c), total bit flip error probabilities for the same three confinement schemes are shown in the presence of the same noise sources  as in plot (b). As expected from the working point selection, all three CNOT designs are bias-preserving with an exponential error suppression factor greater than $2$. In particular, as the gate is fast enough, turning off the confinement of the target cat qubit during the operation does not have a significant impact on the bit-flip suppression. Dotted lines show exponential numerical fits of these bit-flip error probabilities. The exponential suppression factors are given by $\gamma \approx 2.17$ (fully dissipative scheme), $\gamma \approx 2.05$ (combined Kerr), and $\gamma \approx 2.03$ (combined TPE).

\section{Towards Experimental Realization}\label{sec:exp}

The exponential suppression of bit-flips in dissipative cat qubits was experimentally demonstrated in~\cite{lescanne2020exponential} using superconducting circuits as a physical platform. In this experiment the cat qubit resonator was coupled to a low-Q buffer mode with a two-photon exchange Hamiltonian, resulting in an effective two-photon dissipation on the cat qubit resonator. The two-photon exchange Hamiltonian was engineered with an Asymmetrically Threaded SQUID (ATS), a nonlinear coupling element made up of two (ideally) identical junctions shunted by a large inductance. Threading both loops of the ATS at normalized DC flux biases of $0$ and $\pi$ magnetic flux quanta respectively allows to keep only the odd parity mixing terms of the participating modes. Then flux pumping the ATS at frequency $2\omega_a - \omega_b$, where $\omega_{a/b}$ are respectively the cat-qubit and buffer resonator frequencies, creates the required two-photon exchange coupling. Finally, in~\cite{lescanne2020exponential}, not only did the ATS implement the two-photon exchange coupling but it also hosted the low-Q buffer mode.

 \begin{figure}[t!]
         \centering
         \includegraphics[width = 1.0\columnwidth]{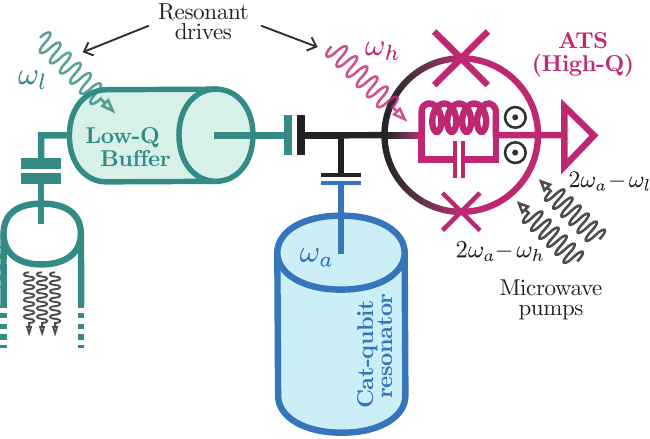}
\caption{\label{fig:circuit-implementation}
Superconducting circuit schematic of a combined TPE and two-photon dissipation confinement. The cat qubit resonator (blue) is capacitively coupled to an ATS (black) that mediates two-photon exchanges both with its self high-Q mode (magenta) and with a low-Q buffer mode (green). The ATS is DC-biased at the normalized flux ($0$,$\pi$) working point and pumped at frequencies $2\omega_a - \omega_h$ and $2\omega_a - \omega_l$, where $\omega_a$, $\omega_h$ and $\omega_l$ are the resonant frequencies of the cat qubit resonator, high-Q buffer and low-Q buffer respectively. To ensure the anharmonicity of the ATS mode (magenta), the ATS junctions are designed to admit slightly different Josephson energies. In addition, both buffers are resonantly driven to control the cat qubit mean number of photons.
}
\end{figure}

The combined TPE and dissipative confinement scheme proposed in this paper can be implemented with minor modifications of this experimental setup. A superconducting circuits scheme of a setup implementing the confinement is represented in Figure~\ref{fig:circuit-implementation}. Similarly to~\cite{lescanne2020exponential}, the cat qubit resonator (in blue) is capacitively coupled to the ATS (in black and magenta) which is DC-biased at the same working point of $0$ and $\pi$. The ATS is further coupled to a linear resonator (in green) which will serve as the low-Q buffer for two-photon dissipation engineering. This leaves the self-mode of the ATS open for other purposes and we use it as the high-Q buffer mode of the device. Thus, instead of having a qubit as the high-Q buffer, this implementation proposes to use a sufficiently anharmonic oscillator mode. In simulations shown later in this Section, we observe that performance degradation remains moderate even with a fully harmonic mode as high-Q buffer, but this option would require further analysis if it were to be considered. Note also that other circuit QED implementations of the TPE Hamiltonian at $\alpha = 0$ were proposed in~\cite{felicetti2018twophoton, felicetti2018ultrastrong}.

To make the high-Q buffer mode anharmonic in this implementation scheme, it is sufficient that the ATS junction energies are designed to be not identical. Note that even in~\cite{lescanne2020exponential} these junctions were not perfectly identical and the amount of asymmetry in that setup might be sufficient to ensure the buffer mode to be anharmonic enough for the design of the present paper. The circuit Hamiltonian engineered with such a setup reads
\begin{equation}
    \begin{split}
        \H =&~\omega_{a} \adag \a + \omega_{h} \bdag_h \b_h + \omega_{l} \bdag_l \b_l \\
        & - 2 E_J \big[ \varepsilon(t) \sin(\hphi) + \eta \widetilde{\mathrm{cos}}(\hphi) \big]
    \end{split}
\end{equation}
where $\a$, $\b_h$ and $\b_l$ are the hybridized modes corresponding to the cat qubit resonator, high-Q and low-Q buffers respectively with $\omega_{a}$, $\omega_{h}$ and $\omega_{l}$ their corresponding resonance frequencies, and where $\widetilde{\mathrm{cos}}(\hphi) = \cos(\hphi) + \hphi^2 / 2$. The total phase across the ATS dipole element reads $\hphi = \varphi_a (\a + \adag) + \varphi_h (\b_h + \bdag_h) + \varphi_l (\b_l + \bdag_l)$. The energy $E_J$ is the average of the two junction energies, $E_J = (E_{J,1} + E_{J,2}) / 2$, and $\eta$ is the asymmetry of the junctions, $\eta = (E_{J,1} - E_{J,2}) / (E_{J,1} + E_{J,2})$. Finally, $\varepsilon(t)$ is an RF flux pumped at two frequencies $2\omega_a - \omega_{h}$ and $2\omega_a - \omega_{l}$ with respective amplitudes $\varepsilon_h$ and $\varepsilon_l$.

\begin{figure*}[t!]
         \centering
         \includegraphics[width = 2.0\columnwidth]{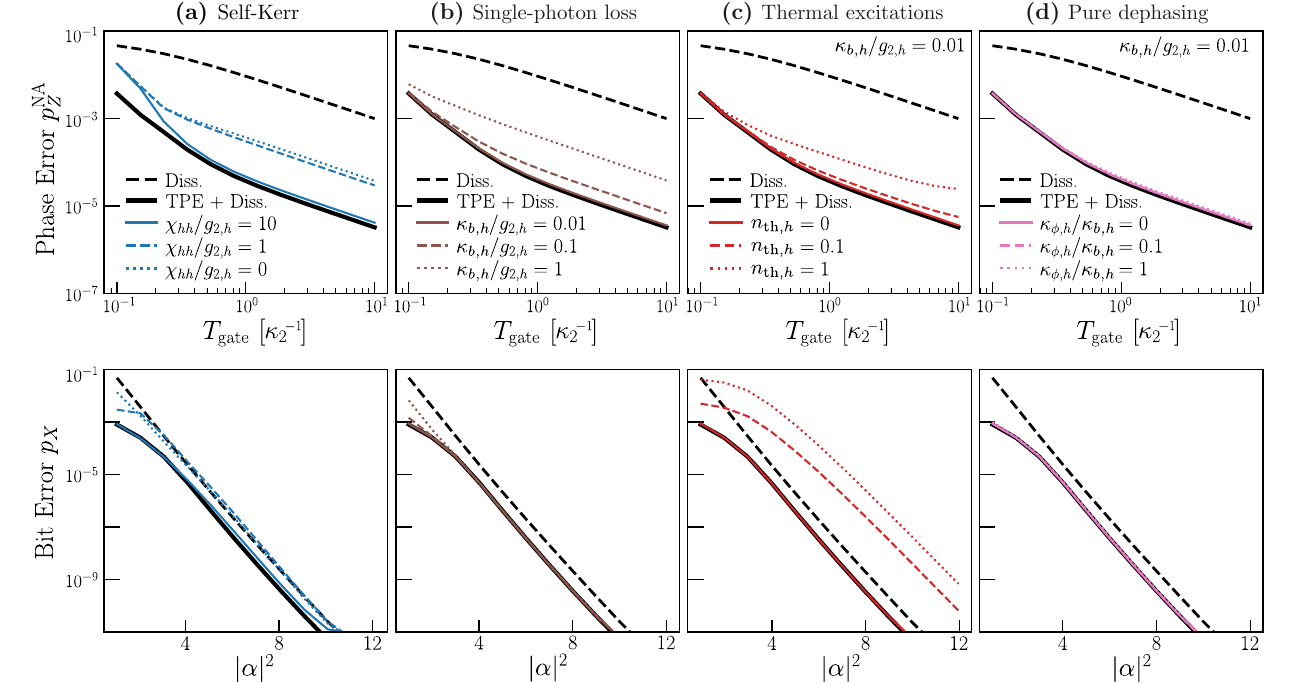}
\vspace{-0.3cm}
\caption{\label{fig:buffer-characterization}
Characterization of the spurious contributions induced by the high-Q buffer mode for a Z gate. Non-adiabatic phase-flip error probability for $|\alpha|^2 = 8$ (top) and bit-flip error probability at $T_\text{gate} = 1/\kappa_2$ (bottom) are shown. In all plots, reference Z gate simulations for two confinement schemes are shown: the fully dissipative confinement scheme (dashed black lines) and the combined TPE and dissipative confinement scheme at the working point $g_{2,h} / \kappa_2 = 10$. Other simulations are the same as for the combined confinement reference simulation, with additional terms as typically resulting from an experimental realization. (a)~The buffer mode is modeled as an anharmonic oscillator, with Hamiltonian $\omega_{b,h} \bdag_h \b_h -\chi_{hh} \Hat{b}_h^{\dagger 2} \b_h^2$ and finite $\chi_{hh}$. (b)~Assuming a perfect two-level buffer, relaxation is considered as a Lindblad term $\kappa_{b,h} \mathcal{D}[\Hat{\sigma}_-]$. (c)~Thermal excitation is added to the the situation of plot (b), thus relaxation is now modeled by the Lindbladian $\kappa_{b,h}(1+n_{\text{th},h}) \mathcal{D}[\Hat\sigma_-] + \kappa_{b,h} n_{\text{th},h} \mathcal{D}[\Hat\sigma_+]$, for a fixed relaxation rate  $\kappa_{b,h} / g_{2,h} = 10^{-2}$.  (d)~Pure dephasing relaxation is added to the situation of plot (b), modeled by a Lindbladian $\kappa_{b,h} \mathcal{D}[\Hat\sigma_-] + \kappa_{\phi, h} \mathcal{D}[\Hat\sigma_z]$, with a fixed single-photon relaxation rate $\kappa_{b,h} / g_{2,h} = 10^{-2}$. Gate times are in units of $1/\kappa_2$.
}
 \end{figure*}

In Appendix~\ref{sec-apdx:circuit-derivation}, we show that this Hamiltonian, after a rotating wave approximation (RWA) and adding resonant drives on both buffer modes, reduces to
\begin{equation}\label{eq:circuit-hamil}
    \begin{split}
        \H =&~g_{2,l} (\a^2 - \alpha^2) \b_l^\dagger + \hc \\
        & + g_{2,h} (\a^2 - \alpha^2) \b_h^\dagger + \hc \\
        & - \chi_{hh} \hat{b}^{\dagger 2}_h \b_h^2 - \chi_{ll} \hat{b}^{\dagger 2}_l \b_l^2 - \chi_{aa} \hat{a}^{\dagger 2} \a^2
        \\
        & - \chi_{ah} \adag \a \bdag_h \b_h - \chi_{al} \adag \a \bdag_l \b_l
        - \chi_{lh} \bdag_l \b_l\bdag_h \b_h
    \end{split}
\end{equation}
where TPE amplitudes are given by $g_{2,x} = E_J \varphi_a^2 \varphi_x \varepsilon_x / 2$, and where self Kerr and cross Kerr terms are given by $\chi_{xx} = \eta E_J \varphi_x^4 / 2$ and $\chi_{xy} = \eta E_J \varphi_x^2 \varphi_y^2$ with $x,y = a$, $h$ or $l$. The parameters can be chosen in such a way that $\chi_{hh}$ strongly dominates all the other coupling terms and that $|g_{2,l}|$ and $|g_{2,h}|$ strongly dominate all the remaining terms. In this manner, the dominant Hamiltonian to be considered is
\begin{equation}\label{eq:circuit-hamil2}
    \begin{split}
        \H_0 =&~g_{2,l} (\a^2 - \alpha^2) \b_l^\dagger +\hc \\
        & + g_{2,h} (\a^2 - \alpha^2) \b_h^\dagger + \hc  \\
        & - \chi_{hh} \hat{b}^{\dagger 2}_h \b_h^2 
    \end{split}
\end{equation}
and all other self-Kerr and cross-Kerr terms can be considered as perturbations to this dynamics. All these perturbation terms are parity preserving but they could in principle induce bit-flip errors on the cat-qubit. However, thanks to the combined  Hamiltonian  and  dissipative confinement, it is sufficient to have the amplitude of these terms small compared to the confinement rate to exponentially suppress their effects. This suppression is facilitated for all the coupling terms, because both buffer modes would nominally be in their ground state for an idling qubit with no further noise.

The fast decay of the low-Q buffer mode at a rate $\kappa_{b,l}$, together with the first term of Eq.~\eqref{eq:circuit-hamil2} yields effectively the  two-photon dissipation $\kappa_2\mathcal{D}[\a^2-\alpha^2]$ on the cat qubit resonator. The effective rate $\kappa_2$ is given by  $4 g_{2,l}^2 / \kappa_{b,l}$ in the limit of $g_{2,l} / \kappa_{b,l}\ll 1$. The second term is the TPE Hamiltonian term. Note that the ratio of TPE to two-photon dissipation confinement rates can actively be tuned through the RF flux amplitudes $\varepsilon_h$ and $\varepsilon_l$. Finally, the third term of Eq.~\eqref{eq:circuit-hamil2} is the Kerr nonlinearity on the buffer mode that is essential to make it anharmonic. This Kerr nonlinearity should be large compared to the TPE Hamiltonian gap, to ensure that terms associated to higher excitations of the high-Q buffer would rotate at a much faster rate than the TPE Hamiltonian contribution, and thus be rejected through RWA. 

In summary, the typical orders of magnitude necessary for the design are given by
\begin{equation}
    \chi_{ah}, \chi_{aa}, \chi_{al} \ll  g_{2,h},g_{2,l} \ll \chi_{hh}
\end{equation}
which, by noting that $\varphi_a,\varphi_l<\varphi_h$, implies
\begin{equation}\label{eq:exp-ineq}
    \eta\varphi_h^2\varphi_a^2\ll \varepsilon_h\varphi_h\varphi_a^2,\varepsilon_l\varphi_l\varphi_a^2\ll\eta\varphi_h^4.
\end{equation}
By design one can choose $\varphi_h$ sufficiently large with respect to $\varphi_l$ and $\varphi_a$, and next tune the junction asymmetry $\eta$ to fulfill the above requirements. An example of a set of experimental parameters that can fulfill these requirements is given by $\varphi_h = 0.2$, $\varphi_a = 0.04$, $\varphi_l = 0.08$, $\eta = 1\%$, $\varepsilon_h = 0.02$, $\varepsilon_l = 0.05$. With these parameters, the condition of Eq.~\eqref{eq:exp-ineq} reads $ 6.4 \cdot 10^{-7} \ll 64 \cdot 10^{-7} \ll 160 \cdot 10^{-7}$. Further assuming an average Josephson energy of $E_J / 2\pi= 80~\mathrm{GHz}$ would yield two-photon exchange rates at $g_{2,l} / 2\pi= g_{2,h}/ 2\pi = 250~\mathrm{kHz}$.

In the rest of this section, we attempt to numerically characterize the impact of spurious Hamiltonian and Lindblad terms related to the high-Q buffer in the TPE confinement. In particular, in Figure~\ref{fig:buffer-characterization}, we investigate a Z gate where we estimate the probability of phase-flip errors at a fixed mean number of photons $|\alpha|^2 = 8$ (top) and bit-flip errors at a fixed gate time $T_\text{gate} = 1/\kappa_2$ (bottom).

The first effect investigated in Figure~\ref{fig:buffer-characterization}(a) is the impact of having an oscillator mode of finite anharmonicity as a buffer, instead of a strict two-level system. Considering a Kerr-type non-linearity for this buffer mode, thus with Hamiltonian $- \chi_{hh} \Hat{b}^{\dagger 2} \b_h^2$, we vary the Kerr strength $\chi_{hh}$ to make the mode more or less anharmonic. This Kerr strength should be large compared to all other frequencies in the system, in order to prevent transitions to higher excited states of the buffer mode. This is demonstrated in panel (a) since only large enough Kerr strengths compared to $g_2$ yield phase errors as good as the reference simulation in solid black lines (where the buffer mode is assumed to be a qubit). Note however that reasonable phase-flip  performances are obtained even with perfectly harmonic buffer modes and that the bit-flip error probability is unaffected, thanks to two-photon dissipation. The option of a harmonic buffer mode would require more careful analysis though, since the corresponding confinement Hamiltonian appears to have a high-dimensional 0-energy eigenspace.

The second effect, investigated in Figure~\ref{fig:buffer-characterization}(b), is high-Q buffer relaxation \ie~finite lifetime of the qubit. This is taken into account in simulations with an additional Lindblad term $\kappa_{b,h} \mathcal{D}[\Hat\sigma_-]$. For large single-photon loss rates $\kappa_{b,h} \gg g_{2,h}$, the high-Q mode in fact becomes low-Q and induces a new two-photon dissipation channel just like $\b_l$, leading to an additional two photon dissipation rate $\tilde\kappa_2 = 4 g_{2,h}^2 / \kappa_{b,h}$. For $\kappa_{b,h} \simeq g_{2,h}$, we can expect the buffer to combine high-Q and low-Q effects. This only shows up as an increase of phase-flip error probabilities during the gate, yet still much better than the performance reached for the fully dissipative gate. In the associated figure, we find that $\kappa_{b,h} / g_{2,h} \leq 0.1$ is enough to obtain performances similar to the reference one.

The third effect investigated, in Figure~\ref{fig:buffer-characterization}(c) is relaxation of the buffer mode in a non-zero temperature environment, corresponding to a Lindbladian $\kappa_{b,h}(1+n_{\text{th},h}) \mathcal{D}[\Hat\sigma_-] + \kappa_{b,h} n_{\text{th},h} \mathcal{D}[\Hat\sigma_+]$. Thermal excitation induces jumps from $\ket{g}$ to $\ket{e}$ which translate to the cat qubit resonator as excitation of the form $(\a^2 - \alpha^{2})^\dag$. Such excitation has a similar effect as direct thermal excitation of the cat-qubit mode already considered in the main text. As far as $g_{2,h}$ is fixed below the working point shown in Fig.~\ref{fig:gamma-bitflip}, the induced leakage is compensated by the two-photon dissipation. Therefore the exponential bit-flip suppression is maintained even though the error probability is increased by a constant factor. This is shown in the bottom plot of Fig.~\ref{fig:buffer-characterization}(c). There only remains an indirect effect on phase-flips due to larger leakage during the gate operation.

Finally, the fourth effect, investigated in Figure~\ref{fig:buffer-characterization}(d), is pure dephasing of the buffer mode, corresponding to a Lindbladian $\kappa_{b,h} \mathcal{D}[\Hat\sigma_-] + \kappa_{\phi, h} \mathcal{D}[\Hat\sigma_z]$. For the values investigated in this figure, \ie~up to $\kappa_{\phi, h} / g_{2,h} = 0.01$, pure dephasing on the buffer mode has no impact on either phase-flip or bit-flip errors during Z gates.

\section{Conclusions}\label{sec:conclusions}
Bosonic encoding and biased-noise qubits have been gaining a lot of attention for their promise of hardware-efficient fault-tolerance. Along these lines, the cat qubits have been investigated with a particular interest and following two confinement approaches: first, a two-photon driven dissipative stabilization scheme and second a two-photon driven Kerr Hamiltonian confinement. In this paper, we have proposed a new confinement design that benefits from the advantages of both dissipative and Hamiltonian approaches. This design, combining the previously known driven dissipative scheme with a Hamiltonian confinement using a Two-Photon Exchange (TPE) Hamiltonian, provides a robust exponential suppression of bit-flip errors with the cat size together with enhanced gate performances, both in speed and fidelity.

Furthermore, this proposal is implementable with only minor modifications of known experimental designs of dissipative cat qubits. We expect that this approach can further reduce the hardware overhead for fault-tolerence with cat qubits. Additional mathematical analysis of these combined confinement schemes and numerical studies of three-qubit Toffoli gates, which are essential for universal quantum computation with repetition cat qubits, will be subject of forthcoming research work.

\begin{acknowledgements}
The authors thank Jeremie Guillaud, Raphael Lescanne, Zaki Leghtas, Lev-Arcady Sellem, Steven Touzard, Nathanael Cottet, Philippe Campagne-Ibarcq, Michiel Burgelman and Pierre Rouchon for  many enlightening discussions. This work was supported by the French Agence Nationale de la Recherche under grant ANR-18-CE47-0005. The numerical simulations were performed using HPC resources at Inria Paris, and using the QuTIP open-source package. The authors thank its developers and maintainers for their work.
\end{acknowledgements}

\appendix
\renewcommand{\appendixname}{APPENDIX}

\section{ENCODING A CAT QUBIT IN AN OSCILLATOR}\label{sec-apdx:catqubitdef}

The codespace of a cat qubit is a two-dimensional subspace spanned by two-component Schr\"{o}dinger cat states~\cite{cochrane1999macroscopically,ralph2003quantum,mirrahimi2014dynamically}. In this paper, the cat qubit logical basis is defined within this codespace as
\begin{equation}\label{eq:catqubitdef}
    \begin{split}
        \ket{0_L} =&~\frac{1}{\sqrt{2}} \left(\ket{\mathcal{C}_\alpha^+} + \ket{\mathcal{C}_\alpha^-} \right) = \ket{\alpha} + \mathcal{O}(e^{-2|\alpha|^2}) \\ 
        \ket{1_L} =&~\frac{1}{\sqrt{2}} \left(\ket{\mathcal{C}_\alpha^+} - \ket{\mathcal{C}_\alpha^-} \right) = \ket{-\alpha} + \mathcal{O}(e^{-2|\alpha|^2})
    \end{split}
\end{equation}
where $\ket{\mathcal{C}_\alpha^\pm}=(\ket{\alpha}\pm\ket{-\alpha}) / \mathcal{N}_{\pm}$ are the even and odd parity cat states, $\mathcal{N}_{\pm} = \sqrt{2 (1 \pm \e^{-2|\alpha|^2})}$ are normalization constants. As such, the cat qubit codespace is a subspace of the infinite dimensional Hilbert space of a quantum harmonic oscillator. Various noise mechanisms could therefore lead to a leakage out of the code space, even though the two-photon dissipation mechanism tends to steer the state back to the codespace.

One way to take into account this leakage in the calculations of the qubit properties (\eg~bit-flip and phase-flip errors) is to specify the qubit state through observables of the quantum harmonic oscillator. Therefore, whole subspaces of the harmonic oscillator state space are associated to a logical qubit value in a way that makes sense compared to the typical measurement process, see later and \cite{guillaud2019repetition}. In other words, the logical qubit value is uniquely defined from Pauli expectation values $\langle \Hat{\sigma}_i \rangle = \Tr[\Hat{J}_i \hrho]$ with $i = x$, $y$, $z$, where $\Hat{J}_i$ are observables to be defined, and where $\Hat{\rho}$ is the density matrix representing the state of the harmonic oscillator system. A natural choice for these observables are the invariants of the two-photon driven dissipative mechanism as defined in~\cite{mirrahimi2014dynamically}. Indeed, we can define $\Hat J_x$ as the photon-number parity operator, whose eigenstates are thus associated to the $\ket{+}$ and $\ket{-}$ logical states, in other words the even or odd parity cats and associated subspaces,
\begin{equation}
    \Hat{J}_x \equiv \Hat{J}_{++} - \Hat{J}_{--}
\end{equation}
where
\begin{subequations}
    \begin{align}
        \Hat{J}_{++} =&~\sum_{n=0} \dm{2n}{2n} \\
        \Hat{J}_{--} =&~\sum_{n=0} \dm{2n+1}{2n+1}.
    \end{align}
\end{subequations}
This observable can be used to compute the phase-flip probabilities of various operations. In a similar manner, the $J_z$ observable can be defined as the invariant
\begin{equation}
    \Hat{J}_z = \Hat{J}_{+-} + \Hat{J}_{+-}^\dagger
\end{equation}
where ($\alpha \in \mathbb{R}$ for simplicity)
\begin{equation}
    \Hat{J}_{+-} = \sqrt{ \frac{2\alpha^2}{\sinh(2\alpha^2)} } \sum_{q=-\infty}^\infty \frac{(-1)^q}{2q + 1} I_q(\alpha^2) \Hat{J}_{+-}^{(q)} 
\end{equation}
and $I_q$ is the modified Bessel function of the first kind, and where
\begin{equation}
    \Hat{J}_{+-}^{(q)} = \left\{ \begin{aligned}
         & \frac{(\adag \a - 1)!!}{(\adag \a + 2q)!!} \Hat{J}_{++} \a^{2q+1} \qquad q \geq 0  \\
         & \Hat{J}_{++} {\Hat{a}}^{\dagger(2|q| - 1)} \frac{(\adag \a)!!}{(\adag \a + 2|q| -1)!!} \quad q < 0
    \end{aligned}\right.
\end{equation}
with $n!! = (n-2)!!\, n$ the double factorial. As shown in~\cite{mirrahimi2014dynamically}, for real $\alpha$, this observable $\Hat{J}_z$ is a very good approximation of $\text{sign}(\a+\adag)$ which signifies if the state has its support in the right or left half plane in the phase space of the harmonic oscillator, thus associated to logical $\ket{0}$ or $\ket{1}$. It is thus a natural observable to evaluate the probability of bit-flips after various operations. Note that $\Hat{J}_z$ swaps the parity, much like a $\sigma_z$ operator swaps the $\ket{+}$ and $\ket{-}$ logical states.


\section{ESTIMATION OF BIT-FLIP ERRORS FOR IDLING CAT QUBITS}\label{sec-apdx:bitflipest}
In this section, we consider an idling cat qubit subject to one of the confinement schemes discussed in the main paper, and derive simplified models for the estimation of bit-flip errors.

\subsection{General Framework}

Let us consider a quantum harmonic oscillator initialized in the pure state 
\begin{equation}\label{eq:initial-state}
    \begin{split}
        \hrho(t=0) =&~\dm{0_L}{0_L} \\
        =&~\dm{\alpha}{\alpha} + \mathcal{O}(\e^{-2|\alpha|^2}).
    \end{split}
\end{equation}
Following the definition of Appendix~\ref{sec-apdx:catqubitdef}, the bit-flip error probability after an idling time $t$ is given by
\begin{equation}
    p_X(t) = \frac{1}{2}\left(1-\Tr[\Hat{J}_z \hrho(t)]\right)
\end{equation}
Computing a bit-flip error probability is therefore equivalent to computing $\hrho(t)$ at all times, and then reinserting it in the above formula. In order to simplify the analytical derivation of $\hrho(t)$, let us neglect the exponential corrections of Eq.~\eqref{eq:initial-state} and assume that the initial state is coherent. Any resulting bit-flip estimation will thus be correct up to this exponential correction. Note that initializing the system in the $\ket{1_L}$ logical state instead would yield identical results, by symmetry. The bit-flip probability simulations of this paper have all been performed using this definition. In the sequel of this section, we will provide the analytical derivation that has been used as a theoretical fit to such bit-flip simulations.

As in Section~\ref{sec:bitflip-confinement}, the system is subject to a cat qubit confinement scheme and to usual decoherence effects such as single-photon relaxation, thermal excitation and pure dephasing,
\begin{equation}\label{eq:s1}
    \begin{split}
        \frac{\dd \hrho}{\dd t} =~g \mathcal{L}_\text{conf} \hrho + \kappa_- \mathcal{D}[\a]\hrho + \kappa_+ \mathcal{D}[\adag]\hrho + \kappa_\phi \mathcal{D}[\adag \a]\hrho
    \end{split}
\end{equation}
where $\mathcal{L}_\text{conf}$ is a confinement superoperator with confinement rate $g$, while $\kappa_- = \kappa_1 (1+ n_\text{th})$ and $\kappa_+ = \kappa_1 n_\text{th}$ are the single-photon relaxation rates, $n_\text{th}$ the average number of thermal photons at the resonator frequency and $\kappa_\phi$ is the pure dephasing noise rate. At all times, the density matrix may be separated in two parts as
\begin{equation}\label{eq:s2}
    \hrho(t) = (1-\varepsilon(t)) \dm{\alpha}{\alpha} + \rhol(t)
\end{equation}
where $\rhol$ represents the part of the density matrix that has leaked out of the coherent subspace $\dm{\alpha}{\alpha}$, such that $\Tr[\rhol(t)] = \varepsilon(t)$. Suppose decoherence rates are small compared to the confinement amplitude, $\kappa_+, \kappa_-, \kappa_\phi \ll g$, and the time of evolution is short compared to the time scale of population leakage out of the coherent state, $\kappa_l  t \ll 1$ where $\kappa_l = \kappa_+ + |\alpha|^2 \kappa_\phi$ is the leakage rate. In this limit, leakage is small and $|\varepsilon(t)| \ll 1$. We first differentiate Eq.~\eqref{eq:s2} with respect to time, and second reinsert Eq.~\eqref{eq:s2} in Eq.~\eqref{eq:s1}, to obtain the two following equations
\begin{equation}\label{eq:drho-1}
    \frac{\dd \hrho}{\dd t} = -\dot{\varepsilon} \dm{\alpha}{\alpha} + \frac{\dd \rhol}{\dd t}
\end{equation}
and
\begin{equation}\label{eq:drho-2}
    \begin{split}
        \frac{\dd \hrho}{\dd t} =&~g \mathcal{L}_\text{conf} \rhol \\
        &+ \kappa_- \mathcal{D}[\a] \dm{\alpha}{\alpha} + \kappa_+ \mathcal{D}[\adag] \dm{\alpha}{\alpha} \\
        &+ \kappa_\phi \mathcal{D}[\adag \a] \dm{\alpha}{\alpha} + \mathcal{O}(\varepsilon^2) \\[1em]
        =&~g \mathcal{L}_\text{conf} \rhol + \kappa_l (\dm{\alpha,1}{\alpha,1} - \dm{\alpha}{\alpha}) \\
        &- \frac{\alpha}{2} (\kappa_1 + \kappa_\phi) \left(\dm{\alpha,1}{\alpha} + \dm{\alpha}{\alpha,1}\right) \\
        &- |\alpha|^2 \kappa_\phi \left( \dm{\alpha, 2}{\alpha} + \dm{\alpha}{\alpha, 2} \right) + \mathcal{O}(\varepsilon^2)
    \end{split}
\end{equation}
where $\ket{\alpha, n} = D(\alpha) (\adag)^n \ket{0}$ is the $n$-th displaced Fock state, and where we have used that $\mathcal{L}_\text{conf} \dm{\alpha}{\alpha} = 0$ by definition. At this point, we project Equations~\eqref{eq:drho-1} and \eqref{eq:drho-2} inside the $\dm{\alpha}{\alpha}$ subspace, which yields
\begin{equation}\label{eq:doteps}
    - \dot{\varepsilon} = g \bra{\alpha} \mathcal{L}_\text{conf} \rhol \ket{\alpha} - \kappa_l.
\end{equation}

For a purely Hamiltonian confinement scheme,
\begin{equation}
    \bra{\alpha} \mathcal{L}_\text{conf} \rhol \ket{\alpha} = -i \bra{\alpha} (\H_\text{conf} \rhol - \rhol \H_\text{conf}) \ket{\alpha} = 0
\end{equation}
since $\H_\text{conf} \ket{\pm \alpha} = 0$ by definition, and thus Eq.~\eqref{eq:doteps} simplifies to $\varepsilon(t) = \kappa_l t$. In other words, for a purely Hamiltonian cat qubit confinement scheme, all leakage out of the codespace due to dissipative noise builds up and never reconverges back to the codespace. Projecting outside of the $\dm{\alpha}{\alpha}$ subspace further yields the master equation on the remainder of the density matrix. For Hamiltonian confinement it just reduces to
\begin{equation}\label{eq:hamil-leak}
    \begin{split}
        \frac{\dd \rhol}{\dd t} =&~-ig [\H_\text{conf}, \rhol] + \kappa_l \dm{\alpha,1}{\alpha,1} \\
        & - \frac{\alpha}{2} (\kappa_1 + \kappa_\phi) \left(\dm{\alpha,1}{\alpha} + \dm{\alpha}{\alpha,1}\right) \\
        & - |\alpha|^2 \kappa_\phi \left( \dm{\alpha, 2}{\alpha} + \dm{\alpha}{\alpha, 2} \right).
    \end{split}
\end{equation}

\subsection{Kerr Confinement}

\begin{figure}[t!]
         \centering
         \includegraphics[width = 1.0\columnwidth]{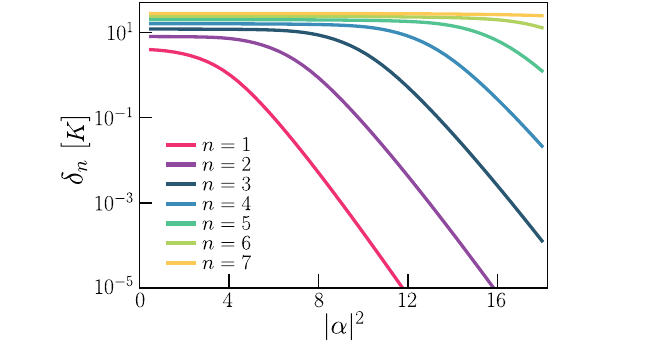}
\vspace{-0.3cm}
\caption{\label{fig-apdx:logdegen}
Energy level spacing between even and odd parity eigenstates of the Kerr Hamiltonian $\delta_n = |e_n^- - e_n^+|$ in units of $K$. Similar plot as in Fig.~\ref{fig:kerr-degeneracies}(a) but with a logarithmic scale on the vertical axis.
}
\end{figure}

We now consider a Kerr Hamiltonian confinement such that $\H_\text{conf} = -\H_\text{Kerr}$, and assume the dynamics of Eq.~\eqref{eq:hamil-leak}. To investigate the associated evolution of $\rhol$, we expand quantum operators into the Kerr Hamiltonian eigenbasis with eigenstates $\{\ket{\phi_n^\pm}\}_{n=0}^\infty$ of energies $\{e_n^\pm\}_{n=0}^\infty$. This eigenbasis has a degenerate ground eigenspace of energy $e_0^\pm = 0$ such that $\ket{\phi_0^\pm} = \ket{\mathcal{C}_\alpha^\pm}$ are the cat qubit basis states. All other eigenstates are separated in two branches of even and odd parities with $0 < e^-_n - e^+_n \ll e^\pm_{n+1} - e^\pm_n$, even though $e^-_n - e^+_n$ grows unbounded in $n$. These energy level spacings $\delta_n = |e_n^- - e_n^+|$ are represented in vertical axis logarithmic scale on Figure.~\ref{fig-apdx:logdegen}. The expansion thus yields
\begin{subequations}
    \begin{align}
        \ket{\alpha} =&~(\mathcal{N}_+ \ket{\phi_0^+} + \mathcal{N}_- \ket{\phi_0^-}) / 2\\
        \label{eq:deflambdanpm} \ket{\alpha,1} =&~\sum_{n>0} \left( \lambda_n^+ \ket{\phi_n^+} + \lambda_n^- \ket{\phi_n^-} \right) \\
        \rhol(t) =&~\sum_{n,m} \sum_{s, r = \pm} \tau_{nm}^{sr}(t) \dm{\phi_n^s}{\phi_m^r}
    \end{align}
\end{subequations}
where $\mathcal{N}_\pm = \sqrt{2 (1 \pm \e^{-2|\alpha|^2})}$ and the $\tau_{nm}^{sr}$ are the variables to be investigated. The $\lambda_n^{\pm}$ are defined by \eqref{eq:deflambdanpm} and they are real for $\alpha$ real.
Reinjecting in Eq.~\eqref{eq:hamil-leak} and projecting onto the $\dm{\phi_n^s}{\phi_m^r}$ density matrix element yields (for $n+m>0$)
\begin{equation}
    \tfrac{\dd}{\dd t}{\tau}_{nm}^{sr} = -i \tau_{nm}^{sr} (e_n^s - e_m^r) + \kappa_{nm}^{sr}
\end{equation}
where $\kappa_{nm}^{sr}$ is the rate at which the corresponding density matrix coefficient is populated, and given by $\kappa_{nn}^{sr} = \kappa_l \lambda_n^s \lambda_n^r$ for terms with $n=m$. We can solve each equation individually for null initial conditions, yielding
\begin{subequations}
    \begin{align}
        \tau_{nn}^{ss}(t) =&~ (\lambda_n^{s})^2\, \kappa_l \, t \\
        \tau_{nm}^{sr}(t) =&~i \kappa_{nm}^{sr} \frac{\e^{-i (e_n^s - e_m^r) t} - 1}{e_n^s - e_m^r} \; .
    \end{align}
\end{subequations}
In this Kerr eigenbasis expansion of $\rhol$, the dominating terms are the diagonal terms $\tau_{nn}^{ss}$ and the terms $\tau_{nn}^{s\bar{s}}$ where $\bar{s} = -s$. All other terms are much smaller due to their $1/(e_n^s - e_m^r)$ dependence and we therefore neglect them. The density matrix $\rhol$ is thus almost block diagonal in the Kerr eigenbasis, with blocks of the form 
\begin{equation}
    \begin{bmatrix}
        \tau_{nn}^{++}(t) & \tau_{nn}^{+-}(t) \\
        \tau_{nn}^{-+}(t) & \tau_{nn}^{--}(t)
    \end{bmatrix}
    = \kappa_l t 
    \begin{bmatrix}
        (\lambda_n^+)^2 & \lambda_n^+ \lambda_n^- \frac{\e^{i \delta_n t} - 1 }{i \delta_n t}\\
        \lambda_n^- \lambda_n^+ \frac{\e^{-i \delta_n t} - 1}{-i\delta_n t} & (\lambda_n^-)^2
    \end{bmatrix}
\end{equation}
where $\delta_n = e_n^- - e_n^+ > 0$. After idling time $T \sim 1 / \delta_n$, coherences of this diagonal block have flipped sign thus inducing bit flip. The bit-flip error is then given by
\begin{equation}\label{eq:kerrbitflip}
    \begin{split}
        p_X(t) =&~\frac{1}{2} \left(1 - \Tr[\Hat{J}_z\hrho]\right) \\
        =&~\frac{1}{2} \left(1 - (1-\kappa_l t) \Tr[\Hat{J}_z \dm{\alpha}{\alpha}] - \Tr[\Hat{J}_z \rhol]\right) \\
        \approx &~\frac{1}{2} \left( \kappa_l t - \Tr[\Hat{J}_z \rhol] \right) \\
        \approx &~\frac{1}{2} \kappa_l t - \frac{1}{2} \sum_{n>0} \left( \tau_{nn}^{-+}(t) \bra{\phi_n^+} \Hat{J}_{+-} \ket{\phi_n^-} + \hc\right) \\
        \approx &~\frac{1}{2} \kappa_l t -  \frac{1}{2} \sum_{n>0} \Big(\tau_{nn}^{+-}(t) + \hc \Big) \\
        \approx &~\frac{1}{2} \kappa_l t \sum_{n>0} \left( (\lambda_n^+)^2 + (\lambda_n^-)^2 - 2 \lambda_n^+ \lambda_n^- \frac{\sin(\delta_n t)}{\delta_n t}\right) \\
        \approx &~\kappa_l t \sum_{n>0} \lambda_n \left(1 - \frac{\sin(\delta_n t)}{\delta_n t} \right)
    \end{split}
\end{equation}
where $\lambda_n = \left[ (\lambda_n^+)^2 + (\lambda_n^-)^2 \right] / 2$ is the value used and shown in the main text. In this derivation, three different assumptions were made. The first is that $\Tr[\Hat{J}_x \dm{\alpha}{\alpha}] = 1$, which is valid up to corrections of order $\exp(-2|\alpha|^2)$. The other two are that $\bra{\phi_n^-} \Hat{J}_{-+} \ket{\phi_n^+} = 1$ and that $(\lambda_n^+)^2 + (\lambda_n^-)^2 = 2 \lambda_n^+ \lambda_n^-$. These two assumptions are true in the limit of large $\alpha$, and would only result in an additional prefactor in front of the sine term which does not change the overall estimation.

\subsection{Two-Photon Exchange Confinement}

The case of a Two-Photon Exchange (TPE) confinement is similar to the one of Kerr confinement. Bit-flip errors are caused by dephasing of the even and odd parity branches of excited eigenstates with respect to each other, which occurs in a time scale given by the eigenstates' energy level spacing. A similar formula for bit-flip error probability can thus be derived by going into the TPE eigenbasis instead of the Kerr one. Those bases are directly related, as explained in the main text. Similar computations then yield
\begin{equation}
    p_X(t) = \kappa_l t \sum_{n \neq 0} \Lambda_n \left[1 - \frac{\sin(\Delta_n t)}{\Delta_n t} \right]
\end{equation}
where $\Lambda_n = \left[ (\Lambda_n^+)^2 + (\Lambda_n^-)^2 \right] / 2$, and where $\Lambda_n^\pm = \braket{\Phi_n^\pm | \alpha, 1}$. From the relation between the two eigenbases, it is not hard to check that $\Lambda_n = \lambda_n/2$

\subsection{Combined Hamiltonian and Dissipative Confinement}
We now consider a harmonic oscillator with a combined Kerr and two-photon dissipation confinement. The master equation that governs the evolution of the system is
\begin{equation}\label{eq:s1-2ph}
    \begin{split}
        \frac{\dd \hrho}{\dd t} =&~i K \left[\H_\text{Kerr}, \hrho \right] + \kappa_2 \mathcal{D}[\a^2 - \alpha^2](\hrho) \\
        &+\kappa_- \mathcal{D}[\a](\hrho) + \kappa_+ \mathcal{D}[\adag](\hrho) + \kappa_\phi \mathcal{D}[\adag \a](\hrho)
    \end{split}
\end{equation}
where $\kappa_2$ is the engineered two-photon dissipation rate and $K$ the engineered Kerr non-linearity. For such a combined scheme, the Kerr confinement induces bit-flip due to the dephasing of even and odd parity eigenstates with respect to each other at a rate given by $\delta_n$. The dissipative confinement on the other hand steers back the state towards the cat qubit manifold at a local rate given by $\kappa_\text{conf} = 4 |\alpha|^2 \kappa_2$, which yields an effective re-convergence timescale given by $t \sim 1 / \kappa_\text{conf}$. Inserting this time scale in Eq.~\eqref{eq:kerrbitflip} yields an estimation of the bit-flip rate. Since two-photon dissipation also induces bit-flip errors in the re-convergence process, an additional bit-flip error rate proportional to $\exp(-2|\alpha|^2)$ must be taken into account, and the total bit-flip error rate is estimated as
\begin{equation}\label{eq:bitflip-combinedkerr}
    \begin{split}
        \Gamma_\text{bit-flip} =&~\kappa_l \exp(-2|\alpha|^2) \\
        &+ \kappa_l \sum_{n > 0} \lambda_n \left[ 1 - \frac{\sin( \delta_n / \kappa_\text{conf})}{\delta_n / \kappa_\text{conf}} \right]
    \end{split}
\end{equation}
For a combined TPE and dissipative confinement scheme, similarly, the total bit-flip error rate is estimated as
\begin{equation}\label{eq:bitflip-combinedtpe}
    \begin{split}
        \Gamma_\text{bit-flip} =&~\kappa_l \exp(-2|\alpha|^2) \\
        &+ \kappa_l \sum_{n \neq 0} \Lambda_n \left[ 1 - \frac{\sin( \Delta_n / \kappa_\text{conf})}{\Delta_n / \kappa_\text{conf}} \right] \; .
    \end{split}
\end{equation}

\section{STEADY STATE OF BIT-FLIP ERRORS AND EXPONENTIAL SUPPRESSION}\label{sec-apdx:gamma}

 \begin{figure}[t!]
         \centering
         \includegraphics[width = 1.0\columnwidth]{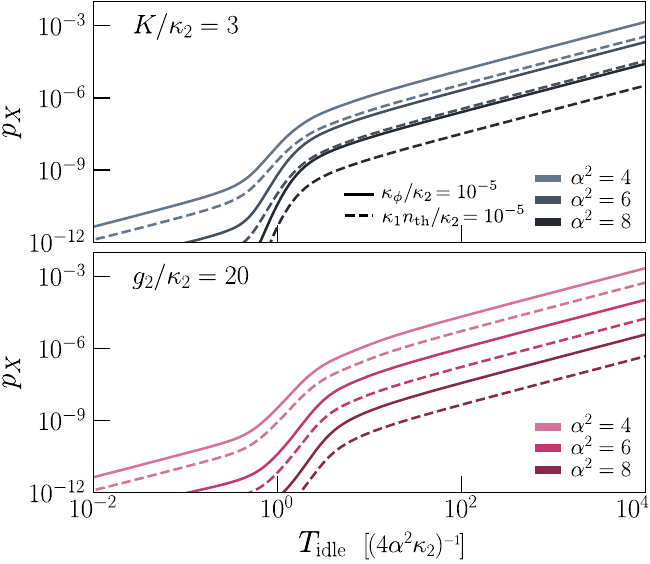}
\vspace{-0.3cm}
\caption{\label{fig-apdx:timesim}
Bit-flip error probabilities of  idling cat qubits confined by combined Kerr (top) and TPE (bottom) schemes at working points $K / \kappa_2 = 3$ and $g_2 / \kappa_2 = 20$. The cat qubit is subject to single-photon loss at rate $\kappa_1 / \kappa_2 = 10^{-3}$, as well as pure dephasing at rate $\kappa_\phi / \kappa_2 = 10^{-5}$ (solid lines) or thermal excitations at rate $\kappa_1 n_\text{th} / \kappa_2 = 10^{-5}$ (dashed lines). The idling time on the horizontal axis is given in units of $1 / \kappa_\text{conf}$ to show a clear transition to the single exponential behavior of the dynamics at around $T_\text{idle} \cdot \kappa_\text{conf} \gtrsim 5$.
}
 \end{figure}

In the paper, bit-flip error rates are computed by fitting an exponential curve $p_X(t) \propto 1 - e^{-\Gamma_\text{bit-flip} t}$ onto numerical simulations of the master equation at large idling times. In the presence of two-photon dissipation, the typical dissipative timescale is $t \sim (\kappa_\text{conf})^{-1} = (4|\alpha|^2 \kappa_2)^{-1}$, and the bit-flip error rate is extracted from simulations at $T_\text{idle} \gg (\kappa_\text{conf})^{-1}$. For purely Hamiltonian simulations, this timescale should in theory be infinite. However, single-photon dissipation also induces re-convergence to the codespace in a typical timescale given by $t \sim (\kappa_1)^{-1}$. In this case, numerical simulations are performed for idling times $T_\text{idle} \gg (\kappa_1)^{-1}$.

Figure~\ref{fig-apdx:timesim} shows bit-flip error probabilities as a function of time for idling cat qubits, under a combination of dissipative confinement respectively with Kerr (top) and TPE (bottom) Hamiltonians. In both plots, a clear exponential behavior of the dynamics is attained after typical idling time $T_\text{idle} \gtrsim 5 (\kappa_\text{conf})^{-1}$, as can be deduced from the linear scaling of bit-flip error probability in the log scale. The slope directly gives $\Gamma_\text{bit-flip}$ for the corresponding value of $\alpha$ and of $K/\kappa_2$ or $g_2/\kappa_2$. 

 \begin{figure}[t!]
         \centering
         \includegraphics[width = 1.0\columnwidth]{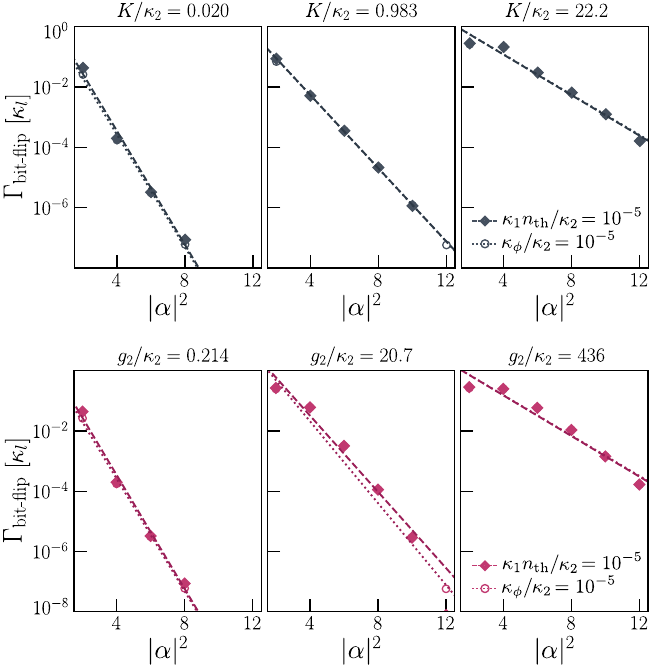}
\vspace{-0.3cm}
\caption{\label{fig-apdx:gammafits}
Bit-flip error rate of an idling cat qubit with a combined Kerr (top) or TPE (bottom) and two-photon dissipation confinement. for increasing values of $K / \kappa_2$ and $g_2 / \kappa_2$. Both simulation data points (markers) and exponential fits of these data points (lines) are shown. The exponential suppression factor $\gamma$ plotted in Fig.~\ref{fig:gamma-bitflip} is extracted from such exponential fits. The cat qubit is subject to single-photon loss at rate $\kappa_1 / \kappa_2 = 10^{-3}$ and either thermal excitations at rate $\kappa_1 n_\text{th} / \kappa_2 = 10^{-5}$ (diamonds) or pure dephasing at rate $\kappa_\phi / \kappa_2 = 10^{-5}$ (circles). Bit-flip error rates are plotted with respect to the leakage rate $\kappa_l = \kappa_1 n_\text{th} + |\alpha|^2 \kappa_\phi$. 
}
 \end{figure}

In Section~\ref{sec:tpe} and Figure~\ref{fig:gamma-bitflip} of the main text, we investigate the exponential reduction of the bit-flip rate as a function of mean photon number in the cat state, by writing $\Gamma_\text{bit-flip} \propto \exp(-\gamma |\alpha|^2)$.  In the limit of a large dissipative to Hamiltonian confinement rate, we observe that $\gamma$ is indeed independent of $\alpha$ and slightly larger than $2$. In the opposite limit of a small dissipative to Hamiltonian confinement rate, it is expected from Eq.~\eqref{eq:bitflip-combinedkerr} and Eq.~\eqref{eq:bitflip-combinedtpe} that $\gamma = 0$ at $|\alpha| = 0$, and that it increases towards an asymptotic limit $\gamma_\infty < 2$ as $|\alpha| \rightarrow \infty$~\cite{putterman2021colored}. In other words, in the limit of a small dissipative to Hamiltonian confinement rate, the exponential dependence of $\Gamma_\text{bit-flip}$ on $|\alpha|^2$ does not hold as such.

We want to estimate the transition from the dissipation-dominated regime, \ie~$\Gamma_\text{bit-flip} \propto \exp(-\gamma |\alpha|^2)$ with $\gamma \geq 2$, to the Hamiltonian-dominated regime, \ie~$\Gamma_\text{bit-flip} \propto \exp(-f(\alpha) \cdot |\alpha|^2)$ with $0 \leq f(\alpha) < 2$.  Figure~\ref{fig:gamma-bitflip} shows the value of $\gamma$ giving the least squares numerical error between the fitting function $h(|\alpha|^2) = -\gamma |\alpha|^2 + c_0$ and the actual function $\ln(\Gamma_\text{bit-flip}) = -f(\alpha) |\alpha|^2 + c_1$, in the range of values $2 \leq |\alpha|^2 \leq 12$. Here $c_0$ and $c_1$ are just optimized proportionality constants. Although the precise values of $\gamma$ depend on this methodology, the transition point between dissipation-dominated and Hamiltonian-dominated regimes would not change with other methodologies.

In Figure~\ref{fig-apdx:gammafits}, several of these numerical fits are shown. In all plots, the bit-flip error rate is plotted against $|\alpha|^2$ with markers showing numerical data, and lines showing the exponential fits with least-squares estimate of $\gamma$. The top plots show the combination of dissipative confinement and Kerr Hamiltonian confinement at increasing values of $K / \kappa_2$, and the bottom plots the combination with TPE Hamiltonian instead. For the smallest values of $K / \kappa_2$ and $g_2 / \kappa_2$ shown, a clear exponential suppression in $|\alpha|^2$ with $\gamma \geq 2$ is found. For the largest values of $K / \kappa_2$ and of $g_2 / \kappa_2$ instead, the data deviates from this exponential dependence, but the linear fit for $\ln(\Gamma_\text{bit-flip})$ still provides a good representation of the bit-flip errors in the range of parameters considered.

\section{ESTIMATION OF GATE-INDUCED PHASE ERRORS FOR Z GATES}\label{sec-apdx:nonadiab-errors}

In this section, we estimate the phase-flip error probability caused by the operation of a Z gate under combined Hamiltonian and dissipative confinement, and in absence of other noise sources. These are also commonly called non-adiabaticity errors. More precisely, we calculate first-order phase errors in the combined Kerr scheme. The calculation for the combined TPE scheme appear to be more complex and require higher order derivations. In this case, we only provide the formula that has been used to analytically fit the numerical results of the main text.

The master equation that describes the Z gate operation proposed in the main text, with a combined Kerr and dissipative confinement scheme, is:
\begin{multline}
    \frac{\dd \hrho}{\dd t} = i K \left[(\a^2 - \alpha^2)^\dagger (\a^2 - \alpha^2), \hrho \right] \\
    - i \varepsilon_Z(t) \left[\a + \adag, \hrho \right] + \kappa_2 \mathcal{D}[\a^2 - \alpha^2]\hrho  \; .
\end{multline}
For the purpose of the following analysis, we apply the Shifted Fock Basis (SFB) method which splits the cat qubit mode into a two-dimensional qubit mode coupled to a harmonic oscillator gauge mode~\cite{chamberland2020building}. While various details on this method can be found in~\cite{chamberland2020building}, here we only need to use that the annihilation operator of the cat-qubit mode $\a$ is well approximated by the operator $\widetilde{\sigma}_z \otimes (\ta + \alpha)$, where $\widetilde{\sigma}_z$ is the Pauli Z operator of the logical qubit described by the cat and $\ta$ is the photon annihilation operator of the gauge mode. The master equation therefore becomes
\begin{equation}
\begin{split}
    \frac{\dd \trho}{\dd t} =&~i K [\tadagd \ta^2 + 2 \alpha^* \tadag \ta^2 + 2 \alpha \tadagd \ta + 4|\alpha|^2 \tadag \ta, \trho] \\[-0.5em]
    &- i \varepsilon_Z(t) \left[\widetilde{\sigma}_z \otimes (\ta + \tadag + 2\Re(\alpha)), \trho \right] \\
    &+ \kappa_2 \mathcal{D}[\ta^2 + 2\alpha \ta]\trho \; .
\end{split}
\end{equation}
Like in \cite{chamberland2020building}, in the limit of large $|\alpha|$, we can keep only the terms of order $|\alpha|^2$ and neglect all terms of order $\alpha$ or $1$. This leading-order approximation is found to be valid for the estimation of phase-flip errors, which are the dominating errors for cat qubit Z gates. Furthermore, we move into the rotating frame with the ideal Z gate Hamiltonian $H_{Z, \text{ideal}} = 2 \Re(\alpha) \varepsilon_Z \widetilde{\sigma}_z$. The master equation, thus describing the remaining error, is then given by:
\begin{equation}\label{eq:sfb-kerr}
    \begin{split}
        \frac{\dd \trho}{\dd t} =&~i \omega_K [\tadag \ta, \trho]  + \kappa_\text{c} \mathcal{D}[\ta]\trho \\
        &-i \varepsilon_Z(t) \left[\widetilde{\sigma}_z \otimes (\ta + \tadag), \trho \right]
    \end{split}
\end{equation}
where $\omega_K = 4|\alpha|^2 K$ and $\kappa_\text{c} = 4 |\alpha|^2 \kappa_2$. From this simplified model, several approaches can be followed to compute the dominant phase errors. More generic and mathematically grounded approaches, such as adiabatic elimination~\cite{azouit2017towards} would lead to the same results.

Here, we write the Langevin equations in the Heisenberg picture for Eq.~\eqref{eq:sfb-kerr}. For the logical qubit part, we only need $\dd \widetilde{\sigma}_z / \dd t = 0$, while for the gauge mode it reads
\begin{equation}
    \frac{\dd \ta}{\dd t} = -i \varepsilon_Z(t) \widetilde{\sigma}_z + i \omega_K \ta - \frac{1}{2} \kappa_\text{c} \ta \; .
\end{equation}
The solution to this equation is given by
\begin{equation}\label{eq:heislan-kerr}
    \begin{split}
        \tilde{a}(t) =&~\tilde{a}(0)\e^{-(\frac{1}{2} \kappa_\text{c} - i \omega_K)t}\\
        &-i \widetilde{\sigma}_z \int_0^t \e^{-(\frac{1}{2} \kappa_\text{c} - i \omega_K) (t - t')} \varepsilon_Z(t') \dd t' \; .
    \end{split}
\end{equation}
The first term vanishes at a rate $\kappa_c$ before the  drive is applied. If $\varepsilon_Z(t)$ varies sufficiently slowly or jumps between constant values on which it stays sufficiently long compared to $\kappa_\text{c}$, then the gauge mode follows the dynamics of the drive up to negligible transients. In this case, Eq.~\eqref{eq:heislan-kerr} simplifies to
\begin{equation}\label{eq:afollows}
    \ta(t) = \frac{i \varepsilon_Z(t)}{i \omega_K - \kappa_\text{c} / 2} \widetilde{\sigma}_z
\end{equation}
and the gauge mode dissipation of Eq.~\eqref{eq:sfb-kerr} results in equivalent $\widetilde{\sigma}_z$ dissipation, \ie
\begin{equation}
    \kappa_\text{c} \mathcal{D}[\ta(t)] = \kappa_\text{c} \frac{\varepsilon_Z(t)^2}{\omega_K^2 + \kappa_\text{c}^2 / 4} \mathcal{D}[\widetilde{\sigma}_z]
\end{equation}
From a more physical viewpoint, any displacement out of the code subspace will result in equivalent dephasing of the qubit due to two-photon dissipation bringing the state back in the cat qubit manifold with a phase error. This effect is mitigated with the additional Kerr Hamiltonian as it limits the displacement amplitude of the gauge mode.

We have thus derived an instantaneous dephasing term on the qubit mode. The total phase-flip error probability on the qubit after a phase gate is thus simply given by the integral of the dephasing rate over the gate time, and reads
\begin{equation}
    p_Z^\text{NA} = \frac{1}{1 + 4K^2 / \kappa_2^2} \frac{\overline{\varepsilon_Z^2} T} {|\alpha|^2 \kappa_2}
\end{equation}
where $\overline{\varepsilon_Z^2} = 1 / T \int_0^T \varepsilon_Z(t)^2 \dd t$ is average squared amplitude of the drive over gate time $T$. This is the same result as in Ref~\cite{chamberland2020building}, with the additional prefactor involving $K / \kappa_2$ which can be used to reduce non-adiabatic phase errors thanks to the Kerr confinement.

Taking into account the single-photon loss of the cat qubit resonator with rate $\kappa_1$, and assuming a constant drive $\varepsilon_Z(t) = \theta / 4 \Re(\alpha) T$ during the gate, thus discarding the short transients in Eq.~\eqref{eq:afollows} when switching the gate Hamiltonian on and off, the total phase error for the Z gate with angle $\theta$ is given by
\begin{equation}\label{eq:kerr-na-pz}
    p_Z = \kappa_1 |\alpha|^2  T + \frac{1}{1 + 4K^2 / \kappa_2^2} \frac{\theta^2}{16 |\alpha|^4 T \kappa_2}
\end{equation}

In the case of the combined TPE confinement, the phase-flip error probability is well fitted with the formula 
\begin{equation}\label{eq:tpe-na-pz}
    p_Z = \kappa_1 |\alpha|^2  T + \frac{1}{1 + 4g_2^2 / \kappa_2^2} \frac{\theta^2}{16 |\alpha|^4 T \kappa_2}+\frac{\theta^2}{32|\alpha|^4T^2g_2^2}. 
\end{equation}
In this formula, the first contribution due to photon loss is the same gate-independent contribution as for all other confinement schemes. The second contribution is similar to the term found for the combined Kerr confinement. The third contribution is however new and specific to the TPE Hamiltonian, where the Z gate Hamiltonian induces hybridization of the codespace with a high-Q buffer state. We leave the derivation of this fitting formula for future work.



\section{CIRCUIT HAMILTONIAN DERIVATION}\label{sec-apdx:circuit-derivation}

 \begin{figure}[t!]
         \centering
         \includegraphics[width = 1.0\columnwidth]{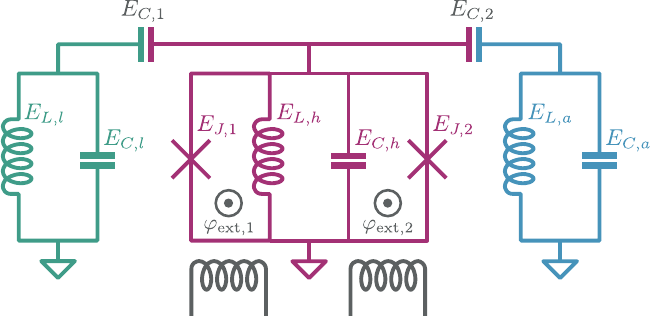}
\vspace{-0.3cm}
\caption{\label{fig:figs3}
Equivalent circuit diagram of the experimental proposition of Fig.~\ref{fig:circuit-implementation}. The cat qubit LC resonator (blue) is capacitively coupled to the ATS circuit element (magenta), itself capacitively coupled to the low-Q buffer LC resonator (green). The high-Q buffer is the self-mode of the ATS, made of two Josephson junctions (nonlinear inductance represented by a cross \& associated capacitor) and an inductance in parallel, and threaded with two flux biases $\varphi_{\text{ext, 1}}$ and $\varphi_{\text{ext, 2}}$. Not shown here: the low-Q buffer resonator is capacitively coupled to a dissipative bath. 
}
 \end{figure}

In this section, we derive the Hamiltonian for the superconducting circuit implementation proposed in Section \ref{sec:exp} of the main paper. Its equivalent circuit is shown in Fig.~\ref{fig:figs3} with an ATS acting as a high-Q buffer, capacitively coupled to the cat qubit resonator and to the buffer resonator. The full Hamiltonian of this circuit reads
\begin{multline}
    \H = \omega_{a,0} \adag \a + \omega_{h,0} \bdag_h \b_h + \omega_{l,0} \bdag_l \b_l \\
    - E_{J,1} \cos(\hphi + \varphi_{\text{ext},1}) - E_{J,2} \cos(\hphi - \varphi_{\text{ext},2})
\end{multline}
where $\a$, $\b_h$ and $\b_l$ are slightly hybridized modes corresponding to the cat qubit, high-Q buffer and low-Q buffer modes respectively, with corresponding mode frequencies $\omega_{a,0}$, $\omega_{h,0}$ and $\omega_{l,0}$. $E_{J,1}$ and $E_{J,2}$ denote the junction energies of the ATS, and $\varphi_{\text{ext},1}$ and $\varphi_{\text{ext},2}$ are the tunable flux biases in each ATS loop as shown in Fig.~\ref{fig:figs3}. The total phase $\hphi$ across the ATS inductance is given by $\hphi = \varphi_a (\adag + \a) + \varphi_h (\bdag_h + \b_h) + \varphi_l (\bdag_l + \b_l)$ where $\varphi_a$, $\varphi_h$ and $\varphi_l$ are the participation ratios of the three modes, deduced from Kirchhoff's laws and the modes hybridization. The corresponding Hamiltonian can be rewritten as 
\begin{equation}
    \begin{split}
        \H =&~\omega_{a,0} \adag \a + \omega_{h,0} \bdag_h \b_h + \omega_{l,0} \bdag_l \b_l \\
        &-2 E_J \big[\cos(\varphi_\Sigma) \cos(\hphi + \varphi_\Delta) \\
        &\qquad +\eta \sin(\varphi_\Sigma) \sin(\hphi + \varphi_\Delta) \big]
    \end{split}
\end{equation}
where $\varphi_\Sigma = (\varphi_{\text{ext}, 1} + \varphi_{\text{ext}, 2}) / 2$ and $\varphi_\Delta = (\varphi_{\text{ext}, 1} - \varphi_{\text{ext}, 2}) / 2$ are the average and difference of the two flux biases, $E_J = (E_{J,1} + E_{J,2}) / 2$ is the average junction energy, and $\eta = (E_{J,1} - E_{J,2}) / (E_{J,1} + E_{J,2})$ is the junction asymmetry. Our working point considers the ATS to be DC biased at the flux bias point defined by
\begin{equation}
    \varphi_\Sigma = \frac{\pi}{2} + \varepsilon(t), \qquad \varphi_\Delta = \frac{\pi}{2}
\end{equation}
with two additional RF fluxes applied on $\varphi_\Sigma$ such that $\varepsilon(t) = \varepsilon_1 \cos(\omega_{p,1} t) + \varepsilon_2 \cos(\omega_{p,2} t)$. This bias point is chosen such that only odd powers of $\hphi$ are turned on for $\eta=0$. The Hamiltonian now reads
\begin{equation}
    \begin{split}
    \H =&~\omega_{a,0} \adag \a + \omega_{h,0} \bdag_h \b_h + \omega_{l,0} \bdag_l \b_l \\
    &-2 E_J \big[ \varepsilon(t) \sin(\hphi) + \eta \cos(\hphi) \big]
    \end{split}
\end{equation}
where we have assumed $|\varepsilon(t)| \ll 1$. By expanding the cosine and sine terms up to fourth order and absorbing the quadratic terms of the cosine in the mode frequencies, the Hamiltonian becomes
\begin{equation}
    \begin{split}
    \H =&~\omega_{a} \adag \a + \omega_{h} \bdag_h \b_h + \omega_{l} \bdag_l \b_l \\
     &-2E_J \big[\varepsilon(t)\hat\varphi-\varepsilon(t) \frac{\hat \varphi^3}{6} + \eta \frac{\hat \varphi^4}{24}  \big].
     \end{split}
\end{equation}

 \begin{figure}[t!]
         \centering
         \includegraphics[width = 1.0\columnwidth]{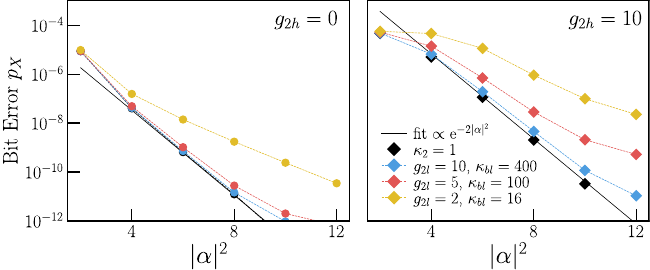}
\caption{\label{fig:figs4}
Bit-flip error probability of an idling qubit for a combined TPE and dissipative confinement with single-photon loss $\kappa_1 = 10^{-3}$ and thermal noise $n_\text{th} = 10^{-2}$. The left panel shows numerical simulations without the TPE Hamiltonian and with a single low-Q buffer mode, while the right panel shows numerical simulations with the TPE Hamiltonian and two buffer modes. In both panels, colored lines show different values of $g_{2,l}$ and $\kappa_{b,l}$ but at fixed effective two-photon dissipation, $\kappa_{2,\text{eff}} = 4 g_{2,l}^2 / \kappa_{b,l}$, computed using the adiabatic elimination formula valid at small $g_{2,l} / \kappa_{b,l}$. Black diamonds show reference simulations with the low-Q buffer mode completely eliminated according to this adiabatic elimination formula, thus actually corresponding to two-photon dissipation $\kappa_2 \mathcal{D}[a^2 - \alpha^2]$. The black line is a numerical fit of the black diamonds with exponential ratio $\exp(-2|\alpha|^2)$.
}
 \end{figure}

We can move to a displaced frame for the three modes to absorb the linear term $-2E_J \varepsilon(t)\hat\varphi$ in the above Hamiltonian. Assuming single-photon decay rates $\kappa_a$, $\kappa_{b,h}$ and $\kappa_{b,l}$, the displacements are given by
\begin{equation}
    \xi_x=\sum_{k=1,2}\frac{-iE_J\varphi_x \varepsilon_k}{i(\omega_x-\omega_{p,k})+\kappa_x/2}e^{-i\omega_{p,k}t},\quad x=a,h,l.
\end{equation}
The Hamiltonian in the displaced frame is given by
\begin{equation}
    \widetilde{H} =\omega_{a} \adag \a + \omega_{h} \bdag_h \b_h + \omega_{l} \bdag_l \b_l +E_J \left[  \varepsilon(t) \frac{\widetilde{\varphi}^3}{3} - \eta \frac{\widetilde{\varphi}^4}{12}\right].
\end{equation}
Here,
\begin{equation}
    \begin{split}
        \widetilde{\varphi} =&~\varphi_a (\adag + \a) + \varphi_h (\bdag_h + \b_h) + \varphi_l (\bdag_l + \b_l) \\
        & + s_1 \e^{-i \omega_{p,1} t} + s_1^* \e^{i \omega_{p,1} t} + s_2 \e^{-i \omega_{p,2} t} + s_2^* \e^{i \omega_{p,2} t}
    \end{split}
\end{equation}
with
\begin{equation}
        s_k = \sum_{x=a,h,l}\frac{i E_J\varepsilon_k\varphi_x^2}{i(\omega_x-\omega_{p,k})+\kappa_x/2},\quad k=1,2.
\end{equation}

Finally, we consider the addition of resonant microwave drives on the buffer modes with amplitudes $\zeta_h(t) = \zeta_h \e^{-i\omega_{h}t}$ and $\zeta_l(t) = \zeta_l \e^{-i\omega_{l}t}$. The Hamiltonian becomes
\begin{equation}
    \begin{split}
        \widetilde{H} =&~\omega_{a} \adag \a + \omega_{h} \bdag_h \b_h + \omega_{l} \bdag_l \b_l \\
        &+[\zeta_h(t) \bdag_h + \hc] + [\zeta_l(t) \bdag_l + \hc] \\
        &+E_J \big[ \varepsilon(t) \frac{\widetilde{\varphi}^3}{3} -  \eta\frac{\widetilde{\varphi}^4}{12} \big]
    \end{split}
\end{equation}
We set the pumping frequencies at $\omega_{p, 1} = 2 \tilde\omega_a - \tilde\omega_h$ and $\omega_{p, 2} = 2\tilde \omega_a - \tilde\omega_l$ where $\tilde\omega_x$ are AC stark shifted frequencies, \ie~including the effects of powers of $\widetilde{\varphi}$ on the actual mode frequencies. By going to the rotating frame of each mode, and performing a Rotating Wave Approximation (RWA), we obtain the effective Hamiltonian
\begin{equation}\label{eq:final-hamil}
    \begin{split}
        \widetilde{H} =&~g_{2,h} (\a^2 - \alpha^2) \b_h^\dagger + \hc \\
        &+g_{2,l} (\a^2 - \alpha^2) \b_l^\dagger + \hc \\
        &-\chi_{hh} (\bdag_h)^2 \b_h^2 - \chi_{ll} (\bdag_l)^2 \b_l^2 - \chi_{aa} (\adag)^2 \a^2
        \\
        &-\chi_{ah} \adag \a \bdag_h \b_h - \chi_{al} \adag \a \bdag_l \b_l
        - \chi_{lh} \bdag_l \b_l\bdag_h \b_h
    \end{split}
\end{equation}\vspace{0.5em}
where the resonant drive amplitudes were set as $\zeta_h = - \alpha^2 g_{2,h}$ and $\zeta_l = - \alpha^2 g_{2,l}$. The various coupling strengths  in this Hamiltonian are given by
\begin{equation}
    \begin{split}
        g_{2,h} =&~E_J \varphi_a^2 \varphi_h \left( \varepsilon_1 / 2  - \eta s_1 \right) \\
        g_{2,l} =&~E_J \varphi_a^2 \varphi_l \left( \varepsilon_2 / 2  - \eta s_2 \right) \\
        \chi_{xx} =&~\eta E_J \varphi_x^4 / 2  \\
        \chi_{xy} =&~\eta E_J \varphi_x^2 \varphi_y^2
    \end{split}
\end{equation}
with $x, y = a, h, l$ and $x \neq y$.

Assuming the Hamiltonian of Eq.~\eqref{eq:final-hamil} with a strong single-photon dissipation on the low-Q buffer mode, $\kappa_{b,l} \mathcal{D}[\b_l]$, it is possible to adiabatically eliminate the buffer mode to obtain an effective two-photon dissipation of amplitude $\kappa_{2,\text{eff}} = 4 g_{2,l}^2 / \kappa_{b,l}$ in the limit of $\kappa_{b,l} \gg g_{2,l}$. Figure~\ref{fig:figs4} showcases a numerical simulation to validate this adiabatic elimination. On the left panel, the usual dissipative cat qubit situation~\cite{lescanne2020exponential,touzard2018coherent} with a single buffer mode is shown. On the right panel, a dissipative cat qubit with the additional TPE confinement at $g_{2,h} / \kappa_2 = 10$, and thus with two buffer modes, is shown. In both panels, a reference simulation is shown in black where the Hamiltonian coupling to the low-Q buffer mode is replaced by the dissipation with $\kappa_{2,\text{eff}}$ resulting from the adiabatic elimination formula. Colored lines then show the results of the full system, including the low-Q buffer mode, for increasing values of $\kappa_{b,l} / g_{2,l}$ at a fixed effective two-photon dissipation rate $\kappa_{2,\text{eff}}$. As can be seen, in both cases the bit-flip error is efficiently suppressed, with a rate tending towards the adiabatic elimination formula once $\kappa_{b,l} / g_{2,l}$ reaches values about $20$ to $40$.

\bibliographystyle{apsrev4-2}
\bibliography{bibliography}

\end{document}